\def\ep{\mbox{\boldmath$\epsilon$}}	% Vector \epsilon
\def\F{{\cal F}}			% Calligraphic F
\def\pd{\partial}			% Partial derivative
\def\td{\mbox{d}}			% Total derivative
\def\<{\left<}				% Species average left bracket
\def\>{\right>}				% Species average right bracket
\def\O{{\cal O}}			% Calligraphic O for order
\def\A{\mbox{\boldmath$A$}}		% Vector A
\def\B{\mbox{\boldmath$B$}}		% Tensor B
\def\C{\mbox{\boldmath$C$}}		% Tensor C
\def\dU{\mbox{\boldmath$U\!$}_1}	% Vector of 1st derivatives of Phi
\def\ddU{\mbox{\boldmath$U\!$}_{11}}	% Tensor of 2nd derivs of Phi
\def\dcU{\mbox{\boldmath$U\!$}_{12}}	% Tensor of cross-derivs of Phi
\def\uuPsi{\mbox{\boldmath$\Psi$}}	% Tensor Psi
\def\vecr{\mbox{\boldmath$r$}}		% Vector r
\def\orig{\mbox{\boldmath$0$}}		% Vector 0 (the origin)
\def\a{{\alpha}}			% alpha for super/subscript
\def\b{{\beta}}				% beta for super/subscript
\def\id{{\rm id}}			% `id' for scripts
\def\ex{{\rm ex}}			% `ex' for scripts
\def\m{{\rm m}}				% `m' for scripts
\def\p{{\rm p}}				% `p' for scripts
\def\ct{\widetilde{\chi}}

\documentstyle[aps,pre,multicol,goodfloat]{revtex}

\input{epsf}
\begin{document}
\draft

\title{Perturbative polydispersity: Phase equilibria of near-monodisperse
systems}
\author{R. M. L. Evans}
\address{Department of Physics and Astronomy, The University of Edinburgh,
	 Mayfield Road, Edinburgh EH9 3JZ, U.K. \\ e-mail: r.m.l.evans@ed.ac.uk}

\date{21 July 2000}
\maketitle

\begin{abstract}
The conditions of multi-phase equilibrium are solved for generic polydisperse
systems. The case of multiple polydispersity is treated, where several
properties (e.g.~size, charge, shape) simultaneously vary from one particle to
another. By developing a perturbative expansion in the width of the
distribution of constituent species, it is possible to calculate the effects of
polydispersity alone, avoiding difficulties associated with the underlying
many-body problem. Explicit formulae are derived in detail, for the
partitioning of species at coexistence and for the shift of phase boundaries
due to polydispersity. `Convective fractionation' is quantified, whereby one
property (e.g.~charge) is partitioned between phases due to a driving force on
another. To demonstrate the ease of use and versatility of the formulae, they
are applied to models of a chemically-polydisperse polymer blend, and of
fluid-fluid coexistence in polydisperse colloid-polymer mixtures. In each case,
the regime of coexistence is shown to be enlarged by polydispersity.
\end{abstract}

%\begin{multicols}{2}

\section{Introduction}

Materials whose constituent elements are much larger than atoms or
simple molecules are now ubiquitous. Such substances include polymer
blends, colloidal suspensions and emulsions. In the manufacture of such large
components, it is impossible to produce a pure strain of truly identical
particles. Instead, each colloidal latex is minutely different in size from its
colleagues, and the number of monomers on each long polymer molecule
inevitably varies across some distribution. These systems are said to be
`polydisperse', though the word is applied only loosely to the polymeric
case, since not every molecule is unique. Some intriguing
experimental \cite{Davesthesis,Bibette91,Koningsveld77} and simulational
\cite{Kofke87,Stapleton90,Bolhuis96,Kofke99,Zhang99} data have been collected
from polydisperse systems. For instance, unusual textures have been found in
polydisperse polymer blends \cite{Clarke95}, and coexisting phases of hard
spheres have been shown to contain unequal proportions of the various
particle sizes, in both simulation \cite{Bolhuis96,Kofke99} and
experiment \cite{Davesthesis,Croatia}.
It is important, then, to understand the statistical
thermodynamics of polydisperse mixtures.

It is well known how to calculate the equilibrium state of a thermodynamic
system \cite{Callen}.
At fixed temperature, an expression must be found for the Helmholtz
free energy. It is then straightforward to determine any property of the
system at equilibrium: its value is that which minimises the free energy.
To establish the concentrations of the system's various constituents, the
minimisation must be performed subject to constraints of global conservation.
The constitution of a uniform system cannot vary, because the amount of each
component is conserved,
but the free energy can sometimes be lowered by partitioning into coexisting
phases of differing compositions.

The non-trivial part of this procedure is to establish a formula for the
Helmholtz free energy. In principle, the subsequent minimisation to derive
phase equilibria is simple.

For a polydisperse system the story is quite different. Here, the number of
components is thermodynamically large. In general, this does not greatly
complicate the (already difficult) many-body problem of formulating
the free energy. However, since that free energy is now a function of
infinitely many conserved variables, the minimisation procedure, though
formally understood \cite{Salacuse82,Gualtieri82}, becomes
intractable.

In the past, many calculations have been performed to study the effects of
polydispersity, using pedagogical model systems with a convenient form of
Hamiltonian \cite{Barrat86,Leroche99},
species distribution \cite{Briano84,Kofke89,Kincaid89}, or approximate
free energy \cite{Gualtieri82,Sollich98,Warren99,Bohle96,Sear99,Bartlett99}.
Also, approximate solutions can be found by making {\it ad hoc} guesses about
the compositions of coexisting phases, and then minimising with respect to just
a few variables such as the overall density of each phase
\cite{Barrat86,Bartlett97}.
Alternatively, the minimisation can be performed numerically, by replacing the
continuous distribution of concentrations with an arbitrary, finite set of
variables \cite{Starling66}.

The above methods invariably involve some way of approximating the many-body
problem inherent in the thermodynamics of the particular system. The present
study differs in that we shall calculate {\em only} the effects of
polydispersity. The analytical procedure was outlined recently
\cite{Evans98,Evans99,Fredrickson98}, and is presented in more detail in the
present article, and applied in new ways to solve a range of problems in the
physics of polydispersity. The method is a perturbative one, whereby
polydisperse phase equilibria are derived from a reference state that is
monodisperse and also in multi-phase equilibrium. We make no assumptions as
to the nature of the system, and require no special properties or
approximations to ensure tractability of the statistical mechanics, since we
shall not address the underlying many-body problem. Instead, we assume that the
many-body properties of the monodisperse reference system are known (whether
from experiment or theory), and derive some exact thermodynamic relations,
applicable to most systems, for the {\em changes} in its properties due to the
introduction of polydispersity.

The aim then is to treat a polydisperse system as a perturbation to a
monodisperse reference state, using the variables that characterise the
polydisperse particles (for
instance, {\em size} and {\em charge}) as expansion parameters. There are two
major obstacles to such a theory. Firstly, the reference state is singular. To
calculate thermodynamic equilibria, a knowledge of the chemical potential of
each species is required. In the reference system, the population is zero
for all species but one. Unfortunately, the chemical
potential of a non-populated species is (as a rule) negative infinity.
The second obstacle is that a narrow distribution of species is
not necessarily a smooth one, and therefore may not be conducive to
small-variable expansion. To proceed, then, we must isolate the badly behaved
functions for exact treatment, and expand only smoothly varying quantities.
The distribution of species is not expanded (contrast \cite{Briano84}), so that
any such distribution may be treated, including a set of delta peaks
representing a finite number of components.
The canonical ensemble is used throughout, as it is of greatest
practical relevance. Thus, the overall mix of species (the `parent
distribution') is specified {\it a priori}, and the pressure and chemical
potentials are derived quantities, used to calculate the resulting
distribution adopted by each coexisting phase after partitioning.

Perturbative methods suggested previously \cite{Kincaid83,Kincaid87} have been
dogged by a proliferation of variables, leading to unwieldy expressions. That
problem is avoided here by relating all properties of the polydisperse system to
just two or three functions that parameterise a generic free energy. An
alternative approach for perturbing about a monodisperse reference is to assume
that all species but one are dilute, and to use their {\em concentrations} as
small expansion
parameters \cite{Gualtieri82}, in the manner of a virial series. In the present
study, by expanding in the width, rather than height, of the distribution of
polydisperse components, we are using a reference state that correctly
treats the many-particle interactions of concentrated, phase-separated
systems. For narrow distributions therefore, this series is expected to
converge more quickly than a virial-like expansion. An exception is at
critical points, where both methods fail.

The formalism and notation are set out in the next section, where a generic
free energy is perturbatively expanded to low order, to derive the pressure
and chemical potentials of a single polydisperse phase. These expressions
are used in section \ref{phaseequilib} to analyse polydisperse phase
equilibria. The normalised distributions, giving the {\em relative} amount of
each species in each phase, are calculated in section \ref{fractsection}.
Their {\em differences} (due to fractionation) are found to obey a very simple
law. For systems that are simultaneously polydisperse in two properties
(e.g.~size and charge), this law is shown to
describe `convective fractionation', whereby one property is partitioned
between phases due to a driving force on the other. In section
\ref{shiftsection}, the effect of polydispersity on the total number
densities at coexistence is calculated. The resulting shift of a phase
boundary along the density-axis of a phase diagram is shown to be
proportional to the variance (standard deviation squared) of the parent
distribution, in the limit of a narrow distribution. Explicit
expressions are given for the cloud- and shadow-point densities, as well as
more general binodals. All the derived formulae quantifying polydisperse
phase equilibria are expressed in terms of a couple of basic parameters of
the free energy. In case the free energy is not known for the polydisperse
system in question, a method for evaluating the relevant parameters is given
in section \ref{pert} for soft interaction potentials, and in section
\ref{hard} for hard spheres and related systems. The method for soft
potentials is extended to determine the lowest-order effect of
polydispersity on correlation functions in appendix \ref{correlations}. To
demonstrate the ease of use of the formulae for phase equilibria, they are
applied in section \ref{FH} to a Flory-Huggins model of chemically
polydisperse polymers, and in section \ref{example2} to fluid-fluid
coexistence in colloid-polymer mixtures, where slight polydispersity
is shown to widen the coexistence region.

\section{Properties of a single phase}
\label{singlephase}

Consider a polydisperse phase of overall number density $\rho$, whose
constituent particles are distinguished by a number $\nu$ of properties
({\it e.g}.~mass, charge, diameter, oblateness\ldots). Each particle is
characterised by an $\nu$-component vector $\ep$, drawn from a narrow
distribution $p(\ep)$ normalised over the whole of $\ep$-space. Let each
component of $\ep$ represent the (dimensionless) fractional deviation of each
property from some reference value. For instance, a hard sphere of radius $r$
in a polydisperse sample would be characterised by a scalar
\mbox{$\epsilon=(r-r_0)/r_0$}, in terms of the reference size $r_0$.
Since the distribution $p(\ep)$ is narrow,
the components of $\ep$ are always small numbers.

To isolate the badly behaved part of the chemical potential, we shall write
the phase's free energy density (which is a functional of the distribution of
species densities $\rho p(\ep)$) in two parts
\begin{equation}
\label{Fsplit}
	\F[\rho p(\ep)] \equiv \F^\id + \F^\ex.
\end{equation}
This is an exact statement, as it serves only to define $\F^\ex$ as the
{\em excess} free energy density, over and above that of a polydisperse ideal
gas of densities $\rho p(\ep)$. The ideal part is simply the sum, over the
continuum of species, of the free energy density of an ideal gas of each
species,
\begin{equation}
\label{Fid}
	\F^\id[\rho p(\ep)] = \int \td^\nu\!\!\epsilon\; \rho p(\ep)\,
	\left\{ \ln(\rho p(\ep)) - 1 \right\}.
\end{equation}
The chemical potentials for the continuum of species are given by a functional
derivative \cite{Salacuse82} of $\F$ with respect to the density of each species,
\begin{equation}
\label{mu}
	\mu(\ep) \equiv \frac{\delta\F}{\delta[\rho p(\ep)]}
\end{equation}
which, from Eq.~(\ref{Fsplit}), splits into two parts,
\begin{equation}
\label{musplit}
	\mu(\ep) = \mu^\id(\ep) + \mu^\ex(\ep).
\end{equation}
By functional differentiation of Eq.~(\ref{Fid}), the ideal part of the chemical
potential is
\begin{equation}
\label{muid}
	\mu^\id(\ep) = \ln (\rho p(\ep))
\end{equation}
which is singular in the monodisperse limit $p(\ep)\to\delta(\ep)$, because it
derives from the entropy of mixing. The
other part, $\mu^\ex(\ep)$, remains well behaved, since it describes the
physics of interactions which vary little from one species to the next.

The obstacles to a small-variable expansion, discussed above, have now been
isolated in the badly-behaved but well-defined function $\mu^\id(\ep)$ in
Eq.~(\ref{muid}), and the narrow, but possibly rapidly varying distribution
$p(\ep)$, also appearing in Eq.~(\ref{muid}). These quantities will be treated
exactly, while the effects of interactions, embodied in the excess free energy
density $\F^\ex$, are expanded in the small variable $\ep$.

\subsection{Expansion of the excess free energy}

The excess free energy density in a phase at equilibrium is a function(al)
$\F^\ex([p(\ep)],\rho)$ of the intensive variables $p(\ep)$ and $\rho$. The
dependence of $\F^\ex$ on temperature and other fields is suppressed in the
notation, and $\F^\ex$ is assumed to be absolutely minimised with respect to
any non-conserved order parameters.

Any distribution $p(\ep)$ is
uniquely defined by its moments $\{\<\ep\>, \<\ep\ep\>,
\<\ep\ep\ep\>, \ldots\}$ which, for vectorial $\ep$, are averages over
$p(\ep)$ of outer products of $\ep$. For instance,
\[
	\<\ep\ep\> \equiv \int \ep\ep\, p(\ep)\, \td^\nu\!\!\epsilon.
\]
Since $p(\ep)$ is narrow, higher moments
are increasingly small. So a controlled small-variable expansion of
$\F^\ex$ is
\begin{equation}
\label{Fex}
	\F^\ex(\rho, \<\ep\>, \<\ep\ep\>, \ldots) =
	\F^\ex(\rho, 0, 0, \ldots) + \<\ep\>\cdot\A(\rho)
	+ \<\ep\ep\> : \B(\rho) + \<\ep\>\<\ep\> : \C(\rho)
	\quad + \O(\epsilon^3)
\end{equation}
where $\F^\ex(\rho, 0, 0, \ldots)$ is the free energy of a phase of
monodisperse particles all characterised by $\ep=\orig$. We shall write
this as $\F^\ex_\m(\rho)$, where the subscript m denotes a quantity in the
monodisperse reference system. The
coefficients $\A$, $\B$ and $\C$ are vector and tensor functions of the
overall density $\rho$. The notation $\O(\epsilon^3)$ indicates terms of
third order in $\ep$ and higher. Equation~(\ref{Fex}) is a generic expansion,
not specific to any particular system. The use of a non-singular expansion,
involving only integer powers of small quantities, has the
status of a conjecture. An alternative motivation for the form of
Eq.~(\ref{Fex}) is given in section \ref{evaluation}.

\subsection{Expansion of the excess chemical potential}

Following Eq.~(\ref{mu}), Eq.~(\ref{Fex}) is differentiated, using the identity
\[
	\frac{\delta\F^\ex}{\delta[\rho p(\ep)]} \equiv \frac{\pd\F^\ex}{\pd\rho}
	+ \frac{\pd\F^\ex}{\pd\<\ep\>}
	\frac{\delta\<\ep\>}{\delta[\rho p(\ep)]}
	+ \frac{\pd\F^\ex}{\pd\<\ep\ep\>}
	\frac{\delta\<\ep\ep\>}{\delta[\rho p(\ep)]} + \ldots
\]
where $\delta\<\ep^m\>/\delta[\rho p(\ep)]=(\ep^m-\<\ep^m\>)/\rho$.
In this notation, $\pd\F^\ex/\pd\<\ep\>$ is a vector whose components are the
derivatives of $\F^\ex$ with respect to each component of $\<\ep\>$.
Similarly, $\pd\F^\ex/\pd\<\ep\ep\>$ is a tensor. Thus we obtain a
general expression, truncated at second order in $\ep$, for the excess
chemical potential $\mu^\ex(\ep)$ of the species with property $\ep$, in a
phase characterised by \mbox{$\{\rho, p(\ep)\}$},
\begin{eqnarray}
\label{muex}
	\mu^\ex(\ep) &=& \mu^\ex_\m(\rho)
	+ \<\ep\> \cdot \big( \A'(\rho)-\A/\rho \big)
	+ \<\ep\ep\> : \big( \B'(\rho)-\B/\rho \big) 
	+ \<\ep\>\<\ep\> : \big( \C'(\rho)-2\C/\rho \big)	\nonumber \\
&& \qquad\qquad
	+ \left. \ep \cdot \big( \A+2\<\ep\>\cdot\C \big)\!\right/\!\rho
	+ \left. \ep\ep : \B\;\right/\!\rho \hspace{5mm} + \O(\epsilon^3)
\end{eqnarray}
where $\mu^\ex_\m(\rho)$ is the excess chemical potential, $d\F^\ex_\m/d\rho$
of a monodisperse reference phase at density $\rho$. Equation~(\ref{muex}) will
be central to the calculations of phase equilibria in section
\ref{phaseequilib}.

\subsection{Expansion of the pressure}

The pressure of a polydisperse phase with density $\rho$ and distribution
$p(\ep)$ can be found from
\[
	P = -\F + \int \mu(\ep)\, \rho\, p(\ep) \: \td^\nu\!\!\epsilon
\]
which is the continuum analogue of the standard thermodynamic relation for
the pressure of a mixture. Using Eqs.~(\ref{Fsplit}) and (\ref{musplit}) it
follows that
\begin{equation}
\label{Prelation}
	P = P^\id - \F^\ex + \rho \int \mu^\ex(\ep)\, p(\ep) \:
		\td^\nu\!\!\epsilon
\end{equation}
where $P^\id$ is the ideal pressure of the polydisperse mixture. It is easily
confirmed from Eqs.~(\ref{Fid}) and (\ref{muid}) that this is the usual ideal gas
pressure $P^\id=\rho$. Substituting the series expressions for the excess
free energy and chemical potential (Eqs.~(\ref{Fex}) and (\ref{muex})) into
Eq.~(\ref{Prelation}), and using the pressure of the monodisperse reference
system at density~$\rho$,
$P_\m(\rho) = \rho - \F^\ex_\m(\rho) + \rho\, \mu^\ex_\m(\rho)$, yields the
expansion for the pressure of the polydisperse phase,
\begin{equation}
\label{P}
	P = P_\m(\rho) + \<\ep\>\cdot \big( \rho\A'(\rho)-\A \big)
	+ \<\ep\ep\>: \big( \rho\B'(\rho)-\B \big)
	+ \<\ep\>\<\ep\>: \big( \rho\C'(\rho)-\C \big)
	\quad + \O(\epsilon^3).
\end{equation}

\section{Phase equilibria}
\label{phaseequilib}

\subsection{Fractionation}
\label{fractsection}

At coexistence between two or more phases, the distribution $p(\ep)$ of
properties is typically different for each phase. We shall assume that the
{\em overall} distribution of all particles in all phases is known (since the
particles might have been synthesised {\it en masse}, before phase separation
was induced). This will be called the parent distribution $p_\p(\ep)$. Without
loss of generality, the origin of $\ep$ will henceforth be set at the mean of
this parent distribution, so that $\<\ep\>_\p\equiv 0$. The aim of
this section is to calculate the distribution $p_\a (\ep)$ in any given
phase $\a$, of $M$ coexisting phases.

The number of particles in phase $\a$ is some fraction $n_\a$ of
the total number in the system. So conservation of material is expressed as
$\sum_\b^M n_\b = 1$.
In fact, a stronger criterion than this holds. The number of particles
belonging to {\em each species} is conserved. That is,
\begin{equation}
\label{conserve}
	\sum_\b^M n_\b\, p_\b(\ep) = p_\p(\ep)
\end{equation}
for all $\ep$.

This is a convenient point at which to introduce a special notation which
will be useful later. If $Q$ is any property of a phase then,
for that phase, let $\widehat{\Delta}[Q]$ be its deviation from the mean
of $Q$ over all phases, weighted by the number of particles in each.
Hence, in phase $\a$,
\[
	\widehat{\Delta}_\a[Q] \equiv Q_\a - \sum_\b^M n_\b Q_\b
	= \sum_\b^M n_\b [Q]^\a_\b.
\]
The notation $[\ldots]^\a_\b$ indicates the difference between the
quantity evaluated in phase $\a$ and in phase $\b$.

Returning to the problem of coexistence, note that, in any two of the $M$
coexisting phases, $\a$ and $\b$ say, the chemical potential is equal,
{\it i.e}.~$\mu_\a(\ep)=\mu_\b(\ep)$ for all $\ep$. Combining
Eqs.~(\ref{musplit}) and (\ref{muid}), this implies
\begin{equation}
\label{mubalance}
	\rho_\a p_\a(\ep)\, \exp \mu_\a^\ex(\ep)
	= \rho_\b p_\b(\ep)\, \exp \mu_\b^\ex(\ep)
\end{equation}
which can be used to eliminate each $p_\b(\ep)$ in Eq.~(\ref{conserve}),
yielding
\begin{equation}
\label{fundamental}
	p_\a(\ep) = \frac{p_\p(\ep)}{\sum_\b n_\b
	\frac{\rho_\a}{\rho_\b} \exp \left[ \mu_\a^\ex(\ep) -
	\mu_\b^\ex(\ep) \right]}.
\end{equation}
We shall now use Eq.~(\ref{muex}) for the excess chemical potential, truncated at
{\em first} order,
\[
	\mu^\ex(\ep) = \mu^\ex_\m(\rho)
	+ \<\ep\>\cdot (\A'-\A/\rho)
	+ \ep\cdot\A/\rho + \O(\epsilon^2)
\]
and expand the exponential in Eq.~(\ref{fundamental}) as
\[
	\exp [\mu^\ex(\ep)]^\a_\b =
	(1+\ep\cdot[\A/\rho]^\a_\b)
	\exp \big[\mu_\m^\ex+\<\ep\>\cdot(\A'-\A/\rho)\big]^\a_\b
	+ \O(\epsilon^2).
\]
So Eq.~(\ref{fundamental}) can be written
\begin{equation}
\label{inbetween}
	p_\a(\ep) = \frac{p_\p(\ep)}{{\cal S}_\a} \: \left(
	1 - \ep \cdot \frac{\sum_\b
	[\A/\rho]^\a_\b n_\b \frac{\rho_\a}{\rho_\b}
	\exp\big[\mu_\m^\ex+\<\ep\>\cdot(\A'-\A/\rho)\big]^\a_\b}{{\cal S}_\a}
	+ \O(\epsilon^2) \right)
\end{equation}
where ${\cal S}_\a \equiv \sum_\b n_\b \frac{\rho_\a}{\rho_\b}
\exp[\mu_\m^\ex+\<\ep\>\cdot(\A'-\A/\rho)]^\a_\b$. If Eq.~(\ref{inbetween}) is
integrated with respect to $\ep$, then the term linear in
$\ep$ disappears, due to the condition $\<\ep\>_\p\equiv \orig$.
We are left with $1=1/{\cal S}_\a + \O(\epsilon^2)$, or
${\cal S}_\a = 1 + \O(\epsilon^2)$, which can be substituted back into
Eq.~(\ref{inbetween}). Finally, integrating Eq.~(\ref{mubalance}) with
$\mu^\ex(\ep)$ expanded to zeroth order yields
$(\rho_\a/\rho_\b) \exp \left[ \mu_\m^\ex \right]^\a_\b
= 1 + \O(\epsilon)$, which we also substitute into Eq.~(\ref{inbetween}). The
result is
\begin{equation}
\label{p}
	p_\a(\ep) = p_\p(\ep) \left\{ 1 - \ep\cdot\widehat{\Delta}_\a[\A(\rho)/\rho]
	+ \O(\epsilon^2) \right\}.
\end{equation}
This is the result we wanted: an expression for the distribution of species
in each phase. Not surprisingly, it is almost the same as the parent
distribution (which remains unapproximated), since all species have similar
properties ($\ep$ is small for each). The lowest-order correction to
$p_\p(\ep)$ is expressed in terms of $\A$, a (vectorial) parameter
of the excess free energy (Eq.~(\ref{Fex})), and $\rho_\b$, the
density of each phase. In principle, the quantities on the right hand side of
Eq.~(\ref{p}) are known. The densities of the phases $\rho_\a$ will differ a
little (by an amount $\delta_\a$, say) from the binodals of the monodisperse
reference system $\rho^\m_\a$, so that $\rho_\a = \rho^\m_\a + \delta_\a$, and
the numbers of particles $n_\a$ will have similar small changes. However,
in Eq.~(\ref{p}) one may substitute {\em either} the monodisperse {\em or} the
polydisperse values of $\rho$ and $n$, whichever are available from theory or
experiment, since the small differences affect only higher-order terms.

A more elegant expression is obtained if we find the {\em difference}
between the normalised distributions $p(\ep)$ of two of the $M$ coexisting
phases. From Eq.~(\ref{p}),
\begin{equation}
\label{Deltap}
	[p(\ep)]^\a_\b \to -p_\p(\ep)\, \ep \cdot \left[ \A/\rho \right]^\a_\b
\end{equation}
which, together with Eq.~(\ref{conserve}), contains the same information as
Eq.~(\ref{p}). The arrow($\to$) indicates the strict asymptotic limit as
$\<\epsilon^2\>_\p \to 0$.

The parameter $\A/\rho$ in Eq.~(\ref{Deltap}), is the lowest-order
coefficient of $\ep$ in Eq.~(\ref{muex}), so we can write
\begin{equation}
\label{gradient}
	[p(\ep)]^\a_\b \to -p_\p(\ep)\, \ep \cdot \left[
      \left.\raisebox{0pt}[0pt][5pt]{\mbox{\boldmath$\nabla_{\!\!\epsilon}$}}
	\mu^\ex(\ep)\right|_{\ep=\orig}\right]^\a_\b.
\end{equation}
A gradient in $\ep$-space is denoted \mbox{\boldmath$\nabla_{\!\!\epsilon}$},
and it is evaluated for phases $\a$, $\b$ in Eq.~(\ref{gradient}) at the mean
species $\ep=\orig$. Equation~(\ref{gradient}) says that, at coexistence, the
distributions separate along the steepest gradient of $\mu^\ex_\a-\mu^\ex_\b$
in \mbox{$\ep$-space}.

It is informative to take moments of Eq.~(\ref{Deltap}), since these will tell
us how the various properties of the particles are correlated within the
phases. One might wish to examine averages of the individual components
$\epsilon_{i}$ of $\ep$, such as $\<\epsilon_{1}\>$ the mean value of
some property such as the size of particles, or
$\<\epsilon_{1}\epsilon_{2}\>$ the cross-correlation between
{\it e.g}.~sizes and charges of the particles. The most general moment, from
which this information can be extracted, is the mean of an outer product of
$m$ vectors $\ep$. That is the $m$th rank tensor
\begin{eqnarray}
\label{momentdef}
   \<\ep^m\> &\equiv& \int\underbrace{\ep\ep\ldots\ep}\,p(\ep)\;
	\td^\nu\!\!\epsilon \\
   & & \qquad\; m	\nonumber
\end{eqnarray}
whose components are all possible correlators of order $m$. From
Eq.~(\ref{Deltap}), the difference of such a moment between two of the
coexisting phases is
\begin{equation}
\label{momentsep}
	\left[\<\ep^m\>\right]^\a_\b = - \<\ep^{m+1}\>_\p \cdot
	\left[\A/\rho\right]^\a_\b + \O(\ep^{m+2})
\end{equation}
which involves the {\em next} moment of the parent distribution, and one
summation (the scalar product) over the $\nu$ polydisperse properties. The
scalar version of Eq.~(\ref{momentsep}) was derived previously for
substances that are polydisperse in a single property, at two-phase
\cite{Evans98} and multi-phase \cite{Evans99} coexistence.

It has been noted \cite{Xu00} that for some distributions, odd moments
$\<\ep^{2k+1}\>_\p$ are not of order $(2k+1)$ in the width of the
distribution but, due to cancellation in the integral (Eq.~(\ref{momentdef})),
are of higher order, so that the RHS of Eq.~(\ref{momentsep}) is zero to
$\O(\ep^{m+1})$ and the unevaluated $\O(\ep^{m+2})$ terms are the dominant
order\footnote{When $m=2$, Eq.~(\protect\ref{momentsep}) relates the second
moment of the daughters to the {\em skew} of the parent
\protect\cite{Evans98}. For highly symmetric distributions the skew is
small (or vanishing), so the LHS of Eq.~(\protect\ref{momentsep}) is small,
though not vanishingly so, as it is limited by the unevaluated
$\O(\ep^{m+2})$ terms. On the other hand, for a strongly asymmetric parent,
by Eq.~(\protect\ref{momentsep}), the daughters have very different second
moments. Thus the skew of the parent is highly influential, contrary to
the impression given in Ref.~\protect\cite{Xu00}.}.
Specifically, this is the case for distributions which become symmetric in
the narrow limit (e.g.~Gaussian, Schultz). For strongly asymmetric
distributions though, excepting unlikely cancellations, the $m$th moment is
of $m$th order in the width. Equations (\ref{Deltap}) and (\ref{gradient})
remain valid for all narrow distributions.

As the notation of Eq.~(\ref{momentsep}) is rather abstract, let us consider a
specific example: a system of particles with a range of radii $r$ and
charges $q$. At equilibrium the particles are partitioned into two
coexisting phases denoted $\alpha$ and $\beta$, and we are interested in the
difference between the average radii in these phases. The overall average
radius throughout the system is called $\<r\>_\p$ and the average charge
$\<q\>_\p$. Applying Eq.~(\ref{momentsep}) with $m=1$,
$\epsilon_{1}=r/\<r\>_\p-1$ and $\epsilon_{2}=q/\<q\>_\p-1$, and taking
the first component gives the answer
\[
	\<r\>_\a-\<r\>_\b = a \left[ \<r^2\>_\p - \<r\>_\p^2 \right] 
	+ b \left[ \<rq\>_\p - \<r\>_\p\<q\>_\p \right]
\]
with coefficients $a=-[A_{1}/\rho]^\a_\b/\<r\>_\p$ and
$b=-[A_{2}/\rho]^\a_\b/\<q\>_\p$. The first term says that, in the absence of
charges, the amount of size fractionation is proportional to the overall
size variance (the {\em square} of the standard deviation), with a
coefficient $a$ indicating the difference between each phase's affinity for
large particles. With charge polydispersity, the second term indicates that
size fractionation can occur even if $a=0$ so that neither phase favours
larger particles. If size and charge are cross-correlated in the parent,
{\it i.e}.~bigger particles tend to have bigger (or smaller) charges, then a
`convective' fractionation of sizes will be driven by $b$, the `driving
force' separating charges. These two contributions are additive for a system
with a narrow distribution.

\subsection{Shift of the phase boundaries}
\label{shiftsection}

So far we have established the lowest order change (with width of the
distribution) of the distribution of properties in each phase, relative to
the original, \mbox{`parent'} distribution (Eq.~(\ref{p})). However, a
knowledge of
the normalised distributions is not a complete characterisation of the
coexisting phases. We would also like to know the overall densities of the
phases, $\rho_\a$. For a monodisperse system, the binodal densities are
independent of the number of particles in each phase. This is not so when
the system is polydisperse, since the size of one phase's share of the
available particles influences the shape of its distribution of species and,
hence, its thermodynamic properties. One coexistence of particular interest
is the `cloud point'. A phase is at its cloud point when its density lies on
the `cloud curve' in the phase diagram. It then coexists with an
infinitesimal amount of another phase, whose density is on the corresponding
`shadow' curve. The cloud point is important from a practical point of view,
since it marks the threshold at which a separation of phases is first
observable, and the resulting morphology and fractionation can have useful
applications. The cloud point also inspires theoretical interest since it is
relatively easy to treat in a two-phase system, where the distribution of
species in the majority phase is simply equal to the parent.

In the present analysis, as in the previous section, we shall calculate the
conditions of general phase equilibria, not just the cloud and shadow curves.
Having found the general solution, we shall return to the special case of
the cloud point.

To establish the phase equilibria, we shall demand that the pressure and
chemical potentials are equal in all phases, using the expressions for
pressure and chemical potential derived in section \ref{singlephase}. 
Before proceeding further, let us take moments of Eq.~(\ref{p}), as this will
lead to some substantial simplifications. We see that the second moment,
$\<\ep\ep\>$, in any given phase is, to lowest order, equal to that in the
parent. However, the mean, $\<\ep\>$, in that phase is also of second order
in the parent's width, since the mean of the parent vanishes by definition.
In orders of the width of the parent,
$\sigma_\p\equiv\sqrt{\<\ep\cdot\ep\>_\p}$, we have
\begin{mathletters}
\label{means}
\begin{eqnarray}
	\<\ep\> &=& -\<\ep\ep\>_\p \cdot \widehat{\Delta}\left[\A/\rho\right] \quad
	+ \O(\sigma_\p^3) \\
	\text{and}\quad \<\ep\ep\> &=& \<\ep\ep\>_\p \quad
	+ \O(\sigma_\p^3).
\end{eqnarray}
So the terms involving $\<\ep\>\<\ep\>$ in Eqs.~(\ref{muex}) and (\ref{P}) can be
dropped, since they are of order $\sigma_\p^4$. Let us clarify this point.
The excess chemical potentials in Eq.~(\ref{muex}) and the pressure in
Eq.~(\ref{P}) were calculated consistently to second order in the deviations
$\ep$ of the properties of a phase's constituent particles from some
reference. In general, then, to evaluate these thermodynamic potentials for
an arbitrary phase, with respect to an arbitrary reference, terms involving
$\<\ep\>\<\ep\>$ must be preserved. However, the choice $\<\ep\>_\p=\orig$,
to fix the arbitrary reference, allows us to drop the terms in Eqs.~(\ref{muex})
and \ref{P} with coefficient $\C$, for coexisting phases. This is because we
have found that when several phases coexist, their mean properties differ
very little from each other, so that \mbox{$\<\ep\>_\a-\<\ep\>_\p$} is
second-order small.
\end{mathletters}

To calculate the densities of the coexisting phases, we first demand that the
pressure $P\!\left(\rho, \<\ep\>, \<\ep\ep\>\right)$ is equal in any two
phases $\a$, $\b$ of the $M$ coexisting phases. We shall write this as
\begin{equation}
\label{Pbalance}
	\left[ \raisebox{0pt}[5pt][5pt]
	  {$P\!\left(\rho, \<\ep\>, \<\ep\ep\>\right)$} \right]^\a_\b = 0
\end{equation}
and use the expression for $P\!\left(\rho, \<\ep\>, \<\ep\ep\>\right)$ in
Eq.~(\ref{P}). Equation~(\ref{Pbalance}) is a condition on $\rho_\a$, the
binodal densities which we want to find. As mentioned above, in the limit of a
narrow parent, these densities are close to the monodisperse binodals
$\rho_\a^\m$, differing by an amount $\delta_\a$, so that
$\rho_\a = \rho^\m_\a + \delta_\a$. We shall determine the small shift
$\delta_\a$. In Eq.~(\ref{P}) we can Taylor-expand the monodisperse pressure
$P_\m(\rho_\a) = P_\m(\rho_\a^\m) + P'_\m(\rho_\a^\m)\,\delta_\a$. It will
transpire that $\delta_\a$ is of order $\sigma_\p^2$, so that it is
sufficient to truncate this Taylor expansion at first order in $\delta_\a$.
The condition of pressure balance (Eq.~(\ref{Pbalance})) becomes
\[
	[ P_\m(\rho_\m) + P'_\m(\rho_\m)\,\delta 
	+\<\ep\>\cdot(\rho\A'-\A)
	+ \<\ep\ep\>:(\rho\B'-\B) ]^\a_\b = 0.
\]

We see that, due to the low order of expansion, the change in density $\delta$
affects
only $P_\m$, the `monodisperse contribution' to the pressure. Effectively,
each of the polydisperse phases can be replaced by a monodisperse phase that
is subject to an `external field'
\mbox{($\<\ep\>\cdot(\rho\A'-\A)+ \<\ep\ep\>:(\rho\B'-\B)$)}.
In this interpretation,
each (fictitious) monodisperse phase responds to the field with a density
change $\delta$, governed by the response function $P'_\m$. The same
interpretation can be applied to the chemical potential balance below.

We can eliminate the first term because the monodisperse reference phases
at densities $\rho_\a^\m$ are in pressure balance, that is
$[P_\m(\rho_\m)]^\a_\b=0$. The density derivative of the pressure in the
monodisperse system, $P'_\m(\rho_\m)$, can be expressed in terms of the
phase's isothermal compressibility,
\begin{equation}
\label{compressibility}
	\kappa \equiv \frac{-1}{V} \left(\frac{\pd V}{\pd P} \right)_{N, T}
	= 1 \left/ \rho\frac{dP}{d\rho} \right.
	= 1 \left/ \rho^2\frac{d\mu}{d\rho} \right. .
\end{equation}
Hence pressure balance dictates
\begin{equation}
\label{sim1}
	\left[ \frac{\delta}{\kappa\rho} 
	+ \<\ep\>\cdot(\A'\rho-\A)
	+\<\ep\ep\>:(\B'\rho-\B) \right]^\a_\b = 0
\end{equation}
for any pair $(\a, \b)$ of the $M$ phases.

Another constraint on the shifts in density $\delta_{\alpha,\beta}$ can be
derived from the equality of all chemical potentials between any pair of
phases, which leads to Eq.~(\ref{mubalance}) for the excess parts. We substitute
from Eq.~(\ref{muex}) for $\mu^\ex(\ep)$, and further substitute
\begin{eqnarray*}
	\mu_\m^\ex(\rho) &=& \mu_\m(\rho) - \ln\rho	\\
	&=& -\ln\rho + \mu_\m(\rho_\m) + \mu'_\m(\rho_\m)\,\delta
\end{eqnarray*}
to yield
\[
	\left[ p(\ep)\,\exp\left\{ \mu_\m(\rho_\m) + \mu'_\m \delta
	\right. \right. + \<\ep\>\cdot(\A'-\A/\rho)
	+ \<\ep\ep\>:(\B'-\B/\rho) + \left. \left. \ep\cdot\A/\rho
	+ \ep\ep :\B/\rho \right\} \right]^\a_\b = 0
\]
Expanding the exponential of small quantities and integrating gives
\[
	[ e^{\mu_\m(\rho_\m)} ( 1 + \mu'_\m \delta
	+ \<\ep\>\cdot\A'
	+ \<\ep\ep\>:\{\B'+\A\,\A/2\rho^2\} ) ]^\a_\b = 0.
\]
Now, applying Eq.~(\ref{compressibility}) and the condition
\mbox{$\left[\mu_\m(\rho_\m)\right]^\a_\b=0$} yields
\begin{equation}
\label{sim2}
	\!\!\!\!\!\left[
	\frac{\delta}{\kappa\rho^2} + \<\ep\>\cdot\A'
	+ \<\ep\ep\>:(\B'+\A\,\A/2\rho^2) \right]^\a_\b \!\! = 0
\end{equation}
which, with Eq.~(\ref{sim1}), forms a pair of simultaneous equations in two
unknowns, $\delta_\a$ and $\delta_\b$.

As an aside, it is interesting that, as well as fixing the $M$ unknowns
$\delta_\a$, the $2(M-1)$ constraints in Eqs.~(\ref{sim1}) and (\ref{sim2})
impose extra conditions on the reference system for $M>2$. The extra
constraints are analogous to the Gibbs phase rule: a single component,
whether monodisperse or near-monodisperse, can exist as a pair of coexisting
phases under a range of conditions, whereas triple or higher coexistence
requires a special choice of external fields \cite{note}. It is
intriguing that the extra conditions imposed on the monodisperse reference
system constrain the values of $\A(\rho)$ and $\B(\rho)$ (defined in
Eq.~(\ref{Fex})), which, though {\em defined} in the monodisperse system, are not
manifestly relevant to its phase equilibria.

It is straightforward to solve the linear equations (\ref{sim1} and
\ref{sim2}) to yield the value of $\delta_\a$, expressed as
\begin{equation}
\label{xa}
	x_\a = \frac{-[y]^\a_\b}{[\rho^{-1}]^\a_\b}
\end{equation}
in terms of
\mbox{\(
	x \equiv \delta/\kappa\rho + \<\ep\>\cdot(\A'\rho-\A)
	+ \<\ep\ep\> : (\B'\rho-\B)
\)}
and of
\mbox{\(
	y \equiv \<\ep\>\cdot\A/\rho
	+ \<\ep\ep\> : (\B/\rho + \A\,\A/2\rho^2),
\)}
where expressions for $\<\ep\>$, $\<\ep\ep\>$ can be substituted
from Eqs.~(\ref{means}). However, the expression is inelegant,
since there are $M$ phases, of which $\b$ is in no way special. The shift
$\delta_\a$ in the $\a$-binodal should have a symmetric dependence on all
other phases. Since $\b$ is arbitrary, we may symmetrize Eq.~(\ref{xa}) by
averaging it over all $\beta$, with respect to any weight factor $z_{\a\b}$:
\[
	x_\a = \frac{- \sum_\b z_{\a\b} [y]^\a_\b / [\rho^{-1}]^\a_\b}
	{\sum_\b z_{\a\b}}
\]
For neatness, we choose $z_{\a\b}\equiv n_\b\,(\rho_\a^{-1}-\rho_\b^{-1})$.

The final answer for the shift $\delta$ of the density of one\footnote{The
subscript $\a$, denoting the phase in question, is now dropped since it
applies to {\em every} quantity in Eq.~(\protect\ref{shift}) (except
$\<\ep\ep\>_\p$ and operands of $\widehat{\Delta}$).}
of the $M$ coexisting phases due to polydispersity is
\begin{equation}
\label{shift}
	\delta = \rho\kappa\, \<\ep\ep\>_\p :
	\left( (\A'\rho-\A)\, \widehat{\Delta}[\A/\rho] - \B'\rho +\B
	+ \frac{\widehat{\Delta}\Big[ \widehat{\Delta}[\A/\rho]\,\widehat{\Delta}[\A/\rho] \Big]
	- 2\widehat{\Delta}\big[\B/\rho\big]}{2\widehat{\Delta}[1/\rho]} \right)
\end{equation}
plus terms of order $\sigma_\p^3$. As this is a lowest-order expression, the
densities $\rho$ appearing on the R.H.S. may be taken as either the
monodisperse or the polydisperse binodal values, whichever is most
convenient.

As yet, we do not know the values of the parameters $\A$, $\B$, $\A'$ and
$\B'$ in Eq.~(\ref{shift}). They will be discussed in Section \ref{evaluation}.
Nevertheless, the structure of Eq.~(\ref{shift}) is informative. As stated
above, $\delta$ is indeed of order $\sigma_\p^2$. If, as is often the case
\cite{Bolhuis96,Kofke99,Bartlett99,Bartlett97,Clarke00,Almarza99}, a phase
diagram is drawn for some system, showing the densities of coexisting phases
against the width $\sigma_\p$ of some parent distribution, then
Eq.~(\ref{shift}) gives the leading-order Taylor expansion of
the shape of the phase boundaries\footnote{This expansion of the equations
of the phase boundaries is in some ways equivalent (though in a quite
different formalism) to the Clausius-Clapeyron-like equation derived by
Bolhuis and Kofke\cite{Bolhuis96} for the $P-\nu$ phase diagram, where $\nu$
parameterises the variance of a Gaussian distribution of activities of
polydisperse hard spheres. The present analysis, though
restricted to narrow distributions, is more general.}.
By definition, they meet the \mbox{($\sigma_\p=0$)-axis} at
the monodisperse values $\rho_\m$, and we have shown that they do so
perpendicularly, since there is no term linear in $\sigma_\p$. According
to Eq.~(\ref{shift}), the {\em curvature}
of each phase boundary is proportional to the isothermal compressibility
$\kappa$ of the phase. Hence the small-variable expansion breaks down at a
critical point, where the compressibility diverges. This is not unexpected,
since the position of the critical point can itself be shifted by the
presence of polydispersity, for instance to a different temperature.
Criticality in the monodisperse system is a coexistence between two (or
more) phases of {\em equal} densities. Under the same conditions, 
slightly-polydisperse phases, being non-critical, will either coexist at
finitely-different densities, or not demix at all. In either case, the
thermodynamic state is far from the monodisperse reference, so the
divergence of Eq.~(\ref{shift}) at criticality is correct.

Equation (\ref{shift}) exhibits a property typical of polydisperse systems. 
For a given parent distribution, the
positions of the phase boundaries depend on the number of particles in each
phase (since this affects the averages in the definition of $\widehat{\Delta}[\ldots]$).
This contrasts with
monodisperse systems, for which tie-lines can be drawn, connecting a {\em set}
of points in the phase diagram's forbidden region to the {\em same} coexisting
end-points. It is interesting that this signature of polydispersity
\cite{Sollich00} is
evident at the lowest order in $\sigma_\p$. Hence, as soon as any shift in
the phase boundaries due to polydispersity is observable in a system, that
shift should vary as a tie-line is traversed.

As mentioned above, a coexisting state of particular interest is that between
an infinitesimal amount of one phase (the `shadow' phase) and a majority phase
defined to be at its `cloud' point. Such a state defines the extreme boundary
of the coexistence region. The movement of the cloud point\footnote{Greek
subscripts were used to denote generic phases, whereas $c$, $s$ indicate
cloud and shadow phases, for which the relative numbers of constituent
particles are specified.}, $\delta_c$ is of
great importance, since this tells an experimenter where, in the phase
diagram, to first expect phase separation to occur. Let the portion of
particles in the shadow phase be the infinitesimally small number $n_s=s$.
Then the fraction belonging to the majority, cloud phase is $n_c=1-s$. For any
property $Q$, $\widehat{\Delta}[Q]$ vanishes in the majority phase as $s\to0$,
since the cloud phase {\em defines} the number-weighted mean. In the cloud
phase,
\[
	\widehat{\Delta}_c[Q] = Q_c - (n_c Q_c + n_s Q_s) = s\, [Q]^c_s
\]
where $[Q]^c_s$ denotes the difference between the values of $Q$ in the cloud
and shadow phases, \mbox{$Q_c-Q_s$}. In the shadow phase,
\[
	\widehat{\Delta}_s[Q] = (1-s)[Q]^s_c.
\]
Substituting these expressions into Eq.~(\ref{shift}) gives the shift of the
cloud point as
\begin{equation}
\label{cloud}
	\delta_c = -\rho_c \kappa_c \<\ep\ep\>_\p : \left( 
	\B'\,_{\!\!\!c}\, \rho_c - \B_c + 
	\frac{[\A/\rho]^c_s[\A/\rho]^c_s+2[\B/\rho]^c_s}{2[1/\rho]^c_s}
	\right).
\end{equation}
The coexisting shadow phase has a density $\rho_s^\m + \delta_s$ where
\begin{equation}
\label{shadow}
	\delta_s = -\rho_s \kappa_s \<\ep\ep\>_\p : \left( 
	\B'\,_{\!\!\!s}\, \rho_s - \B_s + 
	\frac{[\A/\rho]^c_s[\A/\rho]^c_s+2[\B/\rho]^c_s}{2[1/\rho]^c_s}
	+ \left(\A'_s\rho_s-\A_s\right) [\A/\rho]^c_s \right).
\end{equation}

It is not always appropriate to express the position of a phase boundary
in terms of the number density of particles, $\rho=\rho^\m +\delta$. For
instance, the phase diagram is
often represented in terms of the volume fraction of particles, $\phi$. This
is the product of the number density and the volume of a particle. Since
fractionation occurs, so that the mean volume of a particle differs from
phase to phase, $\phi$ and $\rho$ are not equivalent, and the coexistence
region has a different shape in the two representations of the phase
diagram. Quantities such as $\phi$ are easily derived from the above results
as follows. For spherical particles, for instance, using the scalar
$\epsilon$ to denote fractional deviations in
particle radii: $r_i = (1+\epsilon_i) r_\m$, we have
\begin{eqnarray}
\label{phitransform}
	\phi &=& \frac{4}{3}\pi \<r^3 \rho\>
		= \frac{4}{3}\pi r_\m^3 \<(1+\epsilon)^3\>
		(\rho^\m+\delta)	\nonumber  \\
	\Rightarrow \quad \frac{\phi}{\phi_\m} &=& 1 + \frac{\delta}{\rho_\m}
		+ 3\<\epsilon\>+3\<\epsilon^2\> \quad + \O(\sigma_\p^3)
\end{eqnarray}
with the quantities on the RHS given in Eqs.~(\ref{means}) and (\ref{shift}).

In describing the phase diagram, one final quantity of interest is the width
of the coexistence region, sometimes called the
miscibility gap. In terms of $\rho$, this is the range of densities of the
system as a whole, for which it separates into more than one phase. Say the
miscibility gap is bounded by two phases, called $\a$ and $\b$ in the
reference system.
Quenching to one edge of this gap, the cloud point of phase $\a$, will cause
an infinitesimal amount of phase $\b$ to form. At the other extreme, a
vanishing amount of phase $\a$ coexists with majority phase $\b$. So the
miscibility gap is given by application of Eq.~(\ref{cloud}) alone, to
establish the two cloud points of phases $\a$ and $\b$. The miscibility gap
$[\rho_c]^\a_\b$ is then given, in terms of the gap in the monodisperse
reference system, by
\begin{equation}
\label{gap}
	[\rho_c]^\a_\b = [\rho_c^\m]^\a_\b + \frac{\<\ep\ep\>_\p}{[-1/\rho]^\a_\b} : \Big\{
	[\rho\kappa]^\a_\b\, [\B/\rho]^\a_\b + [1/\rho]^\a_\b\,
	[\rho\kappa(\B'\rho-\B)]^\a_\b
	+ [\A/\rho]^\a_\b\, [\A/\rho]^\a_\b\,
	(\rho_\a\kappa_\a + \rho_\b\kappa_\b)/2 \Big\}.
\end{equation}
Note that, with the gap defined to be positive, $[-1/\rho]^\a_\b$ is
positive. In principal, the lowest-order change in the miscibility gap due to
polydispersity, given in Eq.~(\ref{gap}), may be positive or negative, depending
on the system. We shall return to this topic in section \ref{FH}.

\section{Evaluating the parameters}
\label{evaluation}

We have determined the state of thermodynamic equilibrium for polydisperse
phases, in terms of certain properties of the monodisperse reference phases.
Some of these properties, the density $\rho$ and compressibility $\kappa$, are
well established for most common systems. Others, the parameters $\A(\rho)$,
$\A'(\rho)$, $\B(\rho)$ and $\B'(\rho)$, require further discussion. These
parameters appear in the excess free energy, Eq.~(\ref{Fex}), and we require their
values in the monodisperse system (where $\<\ep\> = \<\ep\ep\> = 0$) at some
density $\rho$. One might already have an expression to hand for the free
energy of a particular system, in the form of Eq.~(\ref{Fex}), in which case it
is trivial to identify
the appropriate parameters and substitute them into the various expressions
above. The results for two commonly-studied systems are given in sections
\ref{FH} and \ref{example2}. Alternatively, the required numbers may be
established by experiment, or may require derivation from first principles.

We shall now find some generic expressions for $\A(\rho)$, $\B(\rho)$ and, for
completeness, also $\C(\rho)$, which will inform our physical interpretation of
these quantities. Then, in section \ref{pert}, we shall apply those
expressions, to construct a method for first-principles calculation of the
parameters from thermodynamic perturbation theory. That theory is applicable
whenever the Hamiltonian of the system can be differentiated with respect to
the properties $\ep$ of the particles. Though many systems meet this
criterion, one system of particular theoretical and pedagogical interest, a
set of hard spheres,
does not. Its Hamiltonian is discontinuous, being zero for all physical
configurations and rising discontinuously to infinity if any spheres overlap.
However, hard spheres belong to a special class of systems
(scalable systems) for which the parameter $\A$, which controls the emergence
of fractionation (see Eqs.~(\ref{p}, \ref{Deltap}, \ref{momentsep})) is
calculable by an alternative method, detailed in section \ref{hard}.

To find general expressions for the quantities $\A$, $\B$ and $\C$, let us
Taylor expand a generic excess free energy with respect to the properties
$\ep_i$ of each constituent particle separately, where $i$ labels the $N$
particles. We expand to second order in these $N$ vectorial variables thus
\[
	F^\ex = F^\ex_\m
	+ \sum_{i=1}^N \ep_i\,\cdot \left. \frac{\pd F^\ex}{\pd\ep_i} \right|_\m
	+ \frac{1}{2} \sum_{i,j=1}^N \ep_i \ep_j :
	\left. \frac{\pd^2 F^\ex}{\pd\ep_i \pd\ep_j} \right|_\m + \O(\epsilon^3)
\]
where the subscript m indicates evaluation in the system where
$\ep_i=\orig$ for all $i$. In that system the particles are all the
same, so the derivative $\pd F^\ex/\pd\ep_i$ takes the same vectorial value for
any particle $i$. Hence we may evaluate it for particle number 1, without loss
of generality. Similarly, the second derivate $\pd^2 F^\ex/\pd\ep_i\pd\ep_j$
takes just two tensorial values, depending on whether $i$ and $j$ label the
same particle (number 1, say) or different particles (1 and 2, say). Hence,
dividing the above equation by the system volume $V$ gives the excess free
energy density
\begin{equation}
\label{generic}
	\F^\ex = \F^\ex_\m
	+ \rho \<\ep\> \cdot \left. \frac{\pd F^\ex}{\pd\ep_1} \right|_\m
	+ \frac{1}{2} \rho \<\ep\ep\>
	: \left. \frac{\pd^2 F^\ex}{\pd\ep_1 \pd\ep_1} \right|_\m
	+ \frac{1}{2} \rho \<\ep\>\<\ep\>
	: N \left. \frac{\pd^2 F^\ex}{\pd\ep_1\pd\ep_2} \right|_\m
	+ \O(\epsilon^3).
\end{equation}
Note that, although the final term contains a factor $N$, it remains
intensive, because $\pd^2 F^\ex/\pd\ep_1\pd\ep_2 \sim 1/V$ since particles 1 and
2 interact less as the system grows.

Comparing Eq.~(\ref{generic}) with Eq.~(\ref{Fex}), we can identify expressions
for the coefficients $\A$, $\B$ and $\C$, which we also express in terms of
derivatives of Eqs.~(\ref{Fex}) and (\ref{muex}),
\begin{mathletters}
\label{ABC}
\begin{eqnarray}
	\A = \frac{\pd\F^\ex}{\pd\<\ep\>}
&=&	\rho \frac{\pd F^\ex}{\pd\ep_1}
	= \rho \frac{\td\mu^\ex(\ep)}{\td\ep}	\\
	\B = \frac{\pd\F^\ex}{\pd\<\ep\ep\>}
&=&	\frac{1}{2}\rho \frac{\pd^2 F^\ex}{\pd\ep_1 \pd\ep_1}
	= \frac{1}{2}\rho \frac{\td^2 \mu^\ex(\ep)}{\td\ep\,\td\ep}	\\
	\C = \frac{1}{2} \frac{\pd^2\F^\ex}{\pd\!\<\ep\>\!\pd\!\<\ep\>}
&=&	\frac{1}{2}\rho \,N\! \frac{\pd^2 F^\ex}{\pd\ep_1 \pd\ep_2}
	= \frac{1}{2}\rho\frac{\pd}{\pd\!\<\ep\>}
	\frac{\td\mu^\ex(\ep)}{\td\ep}
\end{eqnarray}
\end{mathletters}
with all formulae {\em evaluated in the monodisperse reference system}. To
clarify the meanings of the various derivatives, let us read Eq.~(\ref{ABC}a)
from left to right. It states that the parameter $\A$, which is a property of
a {\em monodisperse} phase, is the rate of change of excess free energy
{\em density} as the {\em mean} property $\<\ep\>$ of all particles in the
phase is changed. The next equality asserts that $\A$ is the overall number
density $\rho$ times the rate at which the {\em extensive} excess free energy
changes when the property $\ep$ of particle number 1 alone is changed.
Finally from Eq.~(\ref{ABC}a), this is the same as $\rho$ times the variation
in excess chemical potential for different species. (Though different species
are absent from the monodisperse system, their {\em excess} chemical
potentials remain well defined.)

In sections \ref{pert} and \ref{hard} two methods are developed, using the
various relations in Eqs.~(\ref{ABC}) to ascertain the values of $\A$, $\B$ and
$\C$ in any given system.

\subsection{Thermodynamic perturbation theory}
\label{pert}

In this section, polydispersity is treated as a perturbation to the
Hamiltonian of a system. The Hamiltonian of a polydisperse system (neglecting
the trivial kinetic part)
is $H(\Gamma)$, a function of the set $\Gamma$ of particle coordinates (both
positional and internal) which defines a configuration of the system. Given
that a monodisperse reference system in {\em the same} configuration $\Gamma$
has Hamiltonian $H_\m(\Gamma)$, we define the perturbation $\hat{H}(\Gamma)$ by
\begin{equation}
\label{H1}
	H(\Gamma) \equiv H_\m(\Gamma) + \hat{H}(\Gamma).
\end{equation}
By Boltzmann and Gibbs, the Helmholtz free energy of the polydisperse system
is (with units $k_BT\equiv1$)
\[
	F = N\!\int\!\td\ep\;\,p(\ep)\big(\ln[Np(\ep)]-1\big)
	- \ln\int\!\td\Gamma\,\exp[-H(\Gamma)]
\]
and that of the monodisperse system is
\[
	F_\m = N[\ln N -1] - \ln\int\td \Gamma\,\exp[-H_\m(\Gamma)].
\]
In these two equations we may set $H$ and $H_\m$ to zero to find expressions
for the ideal part, which we subtract, leaving the excess part of the free
energies,
\[
	F^\ex = -\ln\int\!\td\Gamma\,\exp[-H(\Gamma)] + \ln\int\!\td\Gamma
\]
and
\[
	F_\m^\ex=-\ln\int\!\td\Gamma\,\exp[-H_\m(\Gamma)]+\ln\int\!\td\Gamma.
\]
By applying Eq.~(\ref{H1}), we see that the difference between these
expressions is
\[
	F^\ex - F_\m^\ex = -\ln \left(
	\frac{\int\!\td\Gamma\,e^{-H_\m}e^{-\hat{H}}}{\int\!\td\Gamma\,e^{-H_\m}}
	\right)
\]
where the argument of the logarithm is the thermal average of
$\exp(-\hat{H}(\Gamma))$ in the monodisperse
system. Let us denote the thermal average of a stochastic variable $\zeta$ by
$\<\!\<\zeta\>\!\>$. This is a Boltzmann-weighted average over configurations,
denoted by double brackets to distinguish it from averages over the
distribution of species $p(\ep)$, for which we have been using single
brackets $\<\ldots\>$. Hence, the excess free energy of a polydisperse
system, in terms of a perturbation $\hat{H}$ to a monodisperse system, is
\begin{equation}
\label{pertF}
	F^\ex = F_\m^\ex - \ln \<\!\!\< e^{-\hat{H}} \>\!\!\>_{\!\m}
\end{equation}
where the subscript m indicates that the thermal average is performed in the
monodisperse system. Equation~(\ref{pertF}) has the familiar form of
thermodynamic perturbation theory but, for polydispersity, it applies
{\em only to the excess part} of the free energy. The ideal parts of the mono-
and polydisperse free energies differ by a non-perturbative amount.

Note that we are not making any approximation of the interactions,
unlike many perturbation theories which use an ideal or harmonic reference
state. Hence, as with our previous analysis, Eq.~(\ref{pertF}) treats the
effects arising purely due to polydispersity, in a fully interacting system.

We now apply the second equality in Eq.~(\ref{ABC}a) to the perturbative
expression for the excess free energy, Eq.~(\ref{pertF}), to find $\A$ by
varying the properties of a single particle in an otherwise monodisperse
system. The first term in Eq.~(\ref{pertF}), $F_\m^\ex$ is a constant and
therefore does not contribute. We find
\[
	\A/\rho = \left.\frac{\pd F^\ex}{\pd\ep_1}\right|_\m
	= \left. -\frac{\pd}{\pd\ep_1}\ln
	\<\!\!\< e^{-\hat{H}} \>\!\!\>_{\!\m} \right|_\m
\]
yielding
\begin{mathletters}
\label{pertABC}
\begin{equation}
\label{A}
	\A/\rho = \<\!\!\!\< \frac{\pd H}{\pd\ep_1} \>\!\!\!\>_{\!\!\m}
\end{equation}
where the `hat' ($\hat{\quad}$) has been dropped from the Hamiltonian $H$
since, by definition, the perturbation $\hat{H}$ is the only part that
varies with $\ep_1$. Similarly, for Eqs.~(\ref{ABC}b) and (\ref{ABC}c),
\begin{eqnarray}
	\frac{2\B}{\rho}
	= \<\!\!\!\< \frac{\pd^2 H}{\pd\ep_1 \pd\ep_1} \>\!\!\!\>_{\!\!\m}
	\!\!+\<\!\!\!\< \frac{\pd H}{\pd\ep_1} \>\!\!\!\>_{\!\!\m}
	\!\!\<\!\!\!\< \frac{\pd H}{\pd\ep_1} \>\!\!\!\>_{\!\!\m}
	\!\!- \<\!\!\!\< 
	\frac{\pd H}{\pd\ep_1}\frac{\pd H}{\pd\ep_1} \>\!\!\!\>_{\!\!\m}
	\hspace{-1cm} \nonumber \\
\label{B} \\
	\frac{2\C}{\rho N}
	= \<\!\!\!\< \frac{\pd^2 H}{\pd\ep_1 \pd\ep_2} \>\!\!\!\>_{\!\!\m}
	\!\!+\<\!\!\!\< \frac{\pd H}{\pd\ep_1} \>\!\!\!\>_{\!\!\m}
	\!\!\<\!\!\!\< \frac{\pd H}{\pd\ep_2} \>\!\!\!\>_{\!\!\m}
	\!\!- \<\!\!\!\< 
	\frac{\pd H}{\pd\ep_1}\frac{\pd H}{\pd\ep_2} \>\!\!\!\>_{\!\!\m}.
	\hspace{-1cm} \nonumber \\
\label{C}
\end{eqnarray}
\end{mathletters}

In principle, the derivatives of the Hamiltonian appearing in Eqs.~(\ref{A}),
(\ref{B}) and (\ref{C}) are known from the microscopic physics of a given system.
For instance, in a system of particles with central, symmetric, pairwise
additive interactions, whose inter-particle potential is
$U(r,\ep_i,\ep_j)$, the Hamiltonian is
\begin{equation}
\label{central}
	H=\frac{1}{2}\sum_{\stackrel{i=1}{ }}^{N}
	\sum_{\stackrel{j=1}{\neq i}}^{N}
	U(|\vecr_i-\vecr_j|,\ep_i,\ep_j).
\end{equation}
In terms of the matrices of derivatives of the potential,
\begin{mathletters}
\label{potlderivs}
\begin{eqnarray}
  \dU(r) &\equiv&\;\left.\frac{\pd U(r,\ep,\orig)}{\pd\ep}\right|_{\ep=\orig}\\
  \ddU(r) &\equiv& \left.\frac{\pd^2 U(r,\ep,\orig)}
	{\pd\ep\,\pd\ep}\right|_{\ep=\orig} \\
  \dcU(r) &\equiv& \left.\frac{\pd^2 U(r,\ep_1,\ep_2)}
    {\pd\ep_1\pd\ep_2}\right|_{\ep_1=\ep_2=\orig}
\end{eqnarray}
this yields
\end{mathletters}
\begin{mathletters}
\label{symmetric}
\begin{equation}
	\A = \rho^2 \int_{0}^{\infty} \dU(r)\, g_\m(r) \; 4\pi r^2\, \td r.
\end{equation}
Here, $g_\m(r)=\<\!\<\rho(\orig)\rho(\vecr)\>\!\>_\m/\rho^2$ is the radial
distribution function in the monodisperse reference phase. To obtain
Eq.~(\ref{symmetric}a), we have used the fact that the density of particle centres
in the system at any instant is
$\rho(\vecr) = \sum_{i=1}^N \delta^{(3)}(\vecr-\vecr_i)$.
Equations~(\ref{B}, \ref{C}) can similarly be evaluated for central, symmetric,
pairwise additive interactions, giving
\begin{eqnarray}
  \B &=& \frac{\rho^2}{2} \int^{\infty}_0 \left[ \ddU(r)-\dU(r)\dU(r) \right]
	g_\m(r)\,4\pi r^2\,\td r
	+ \frac{\rho^3}{2} \int \td^3 r\, \td^3 r'\: \dU(r)\dU(r') \left[
	g_\m(r)\,g_\m(r') - g^{(3)}_\m(\vecr,\vecr')	\right]	\\
  \C &=& \frac{\rho^2}{2} \int^{\infty}_0 \left[ \dcU(r)-\dU(r)\dU(r) \right]
	g_\m(r)\,4\pi r^2\,\td r
	- \frac{3\rho^3}{2} \int \td^3 r\, \td^3 r'\: \dU(r)\dU(r')
	g^{(3)}_\m(\vecr,\vecr')	\nonumber  \\
  & &	+ \; \frac{\rho^4}{2} \int\td^3r\,\td^3r'\,\td^3r''\: 
	\dU(r)\dU(\left|\vecr'-\vecr''\right|)
	\left[ g_\m(r)\,g_\m(\left|\vecr'-\vecr''\right|)
	- g^{(4)}_\m( \vecr,\vecr',\vecr'') \right]
\end{eqnarray}
where $g^{(3)}_\m(\vecr,\vecr')\equiv
\<\!\<\rho(\orig)\rho(\vecr)\rho(\vecr')\>\!\>_\m/\rho^3$
and $g^{(4)}_\m(\vecr,\vecr',\vecr'')\equiv
\<\!\<\rho(\orig)\rho(\vecr)\rho(\vecr')\rho(\vecr'')\>\!\>_\m/\rho^4$.
Notice that Eqs.~(\ref{symmetric}a,~b,~c) require a knowledge of spatial
correlations in the monodisperse system only. In the appendix, it is shown how
general thermal averages are perturbed by polydispersity, for such systems
with soft, pairwise interactions.
\end{mathletters}

Equations~(\ref{pertABC}) and~(\ref{symmetric}) allow the parameters affecting
phase equilibria (to lowest order in the width of the distribution) to be
calculated for any system with soft (i.e.~differentiable) interactions.
We shall put these expressions to use on some real systems in
sections~\ref{FH} and~\ref{example2}.

\subsection{Scalable systems: The special case of hard spheres}
\label{hard}

In this section we shall consider a system of prevalent theoretical
interest, polydisperse hard spheres. Hard spheres interact via a potential
which is zero, except for configurations where the spheres overlap, which are
forbidden and hence have infinite potential energy. Hard-sphere systems
attract interest because they exemplify substances with short-range repulsive
interactions, while lacking a characteristic energy scale. This leads to
temperature-independent behaviour which is purely entropy-driven, and engenders
simplicity due to the small number of tunable parameters. Unfortunately, hard
spheres are an exceptional case to which the thermodynamic perturbation theory
in section \ref{pert} cannot be applied, since the internal energy is zero and
the interaction potential is discontinuous. Hence the values of the parameters
$\A$, $\B$ and $\C$ cannot be found by that method, though their formulations
in Eqs.~(\ref{ABC}) still hold.

In this section we shall exploit a special property of the hard-sphere system
to establish the exact value of the parameter $\A$ (previously
approximated\cite{mistake}). In fact the method can be
applied to any system of particles whose interactions are `scalable' --- a
term which will be elucidated below. With a knowledge of $\A$, the
partitioning of hard spheres between slightly-polydisperse phases is fully
determined in Eq.~(\ref{p}), (\ref{Deltap}) or (\ref{momentsep}). Unfortunately it
is not clear how such a method might be used to determine $\B$, the other
parameter on which the phase boundaries depend. In section~\ref{example2} its
value is taken from an approximate expression for the free
energy of polydisperse hard spheres, of which there are many examples in the
literature\cite{Boublik70,Barrat86,Bartlett99}.

To calculate $\A$, the coefficient of $\<\ep\>$ in the free energy expansion
(Eq.~(\ref{Fex})), for polydisperse hard spheres, we use the first equality of
Eq.~(\ref{ABC}a). Let the radius $r$ of each sphere be measured relative to a
reference length $r_0$ thus: \mbox{$r=(1+\epsilon)\,r_0$}. As this is the
only property that varies from sphere to sphere, the value $\epsilon$ which
characterises the particles is a scalar, so Eq.~(\ref{ABC}a) reduces to a scalar
equation. That equation states that $A$ is the rate at which the free energy
density changes when the first moment $\<\epsilon\>$ of size deviations is
increased, while the other moments remain stationary\footnote{In a monodisperse
system, the distribution is a delta function. It is therefore not possible to
vary the mean while holding the other moments {\em constant}. Nevertheless,
being of higher order in the small quantity $\ep$, the higher moments are
{\em stationary}, so the partial derivative is well defined.}. When this
hypothetical change is made, the particles all grow by the same amount, so the
system remains monodisperse. In terms of the unique radius $r$ of the
monodisperse particles, the derivative is
\[
	\frac{\pd\F^\ex}{\pd\<\epsilon\>} = r_0 \frac{\pd\F^\ex}{\pd r}
\]
which will be evaluated at $r=r_0$. One can imagine making this change in two
stages: firstly the whole system is scaled up, so that the particles, and the
space between them, and the volume of the system all increase, while the
concentration $\phi$ (the fraction of space occupied by particles) is held
constant. Then the system is compressed back to its original volume while the
particles remain at their new large size, so the concentration increases. The
resulting change in the extensive excess free energy ($F^\ex=V\F^\ex$) is
given by
\[
	\left(\frac{\pd F^\ex}{\pd r}\right)_{\!V,N}
	= \left(\frac{\pd F^\ex}{\pd r}\right)_{\!\phi,N}
	- \left(\frac{\pd F^\ex}{\pd V}\right)_{\!r,N}
	\, \left(\frac{\pd V}{\pd r}\right)_{\!\phi,N}.
\]
The last term here is due to the reversible work done against excess pressure
when the system is compressed, with $(\pd V/\pd r)_{\phi,N}=3V/r$ being a
conversion factor between length and volume. So we have
\[
	\left(\frac{\pd F^\ex}{\pd r}\right)_{\!V,N}
	= \left(\frac{\pd F^\ex}{\pd r}\right)_{\!\phi,N}
	+ 3\frac{P^\ex V}{r}.
\]
The first term on the right hand side is
trivial to calculate because there exist no length scales (such as the
range of an interaction) that remain fixed as the particles grow. Hence the
growth is simply an overall scaling\cite{Kofke89}. This is what was meant by
the term
``scalable system" above. By writing the free energy in terms of the usual
configurational integral over particle positions measured in units of the
thermal de Broglie wavelength, it is easy to establish that this term
vanishes, since the excess part does not vary with the overall scale factor.
Hence, for hard spheres\cite{equivalent},
\begin{equation}
\label{AHS}
	A = 3P^\ex = 3(P-\rho)
\end{equation}
where, as usual, $k_{B}T$ has been set to unity. Since, at equilibrium, $P$
is a constant for all phases, the coefficient in
Eqs.~(\ref{Deltap},~\ref{momentsep}) may be written
$[A/\rho]^{\alpha}_{\beta} = 3P[\rho^{-1}]^{\alpha}_{\beta}$
for size-polydispersity in scalable systems.

Though it is not apparent that an analogous method exists for calculating
$B$, the parameter $C$ is calculable from the first equality
of Eq.~(\ref{ABC}c), by taking the second derivative of excess free energy
density with respect to particle size in a monodisperse scalable system.
We shall not perform that calculation here.

For monodisperse hard spheres, there is a transition \cite{Hoover68,Russel89}
from a crystal of volume
fraction 0.545 to a fluid phase of volume fraction 0.494 which, according
to the Carnahan-Starling equation of state \cite{Carnahan69},
exists at a pressure $6.17\pm0.02$,
in units scaled by $k_B T$ and by the volume of a particle. Hence, from
Eq.~(\ref{AHS}), the coefficient $[A/\rho]^{\rm crystal}_{\rm fluid}$, which
appears in Eq.~(\ref{momentsep}), evaluates to $3.51\pm0.04$. This is consistent
with simulation results in which the average sizes of polydisperse hard spheres
were measured in coexisting fluid and crystal phases, and the difference
plotted against the variance of the overall distribution\cite{Bolhuispriv}.
The gradient was found to be $3.55\pm0.01$, and changed by only a few percent
up to polydispersities of 6\% and more.

\section{Example: Chemically polydisperse polymers}
\label{FH}

It has been necessary to plough through a fair amount of mathematics in
order to arrive at some simple and practical formulae. The convenience of
this approach is that the lengthy derivations have been performed once and
for all. They will not require repetition for each application. Lest the
reader has lost sight of the wood for the trees, let us demonstrate how
easily the formulae may be applied to a model of polymeric phase
equilibria.

\begin{figure} \displaywidth\columnwidth
  \epsfxsize=8.5cm
  \begin{center}
  \leavevmode\epsffile{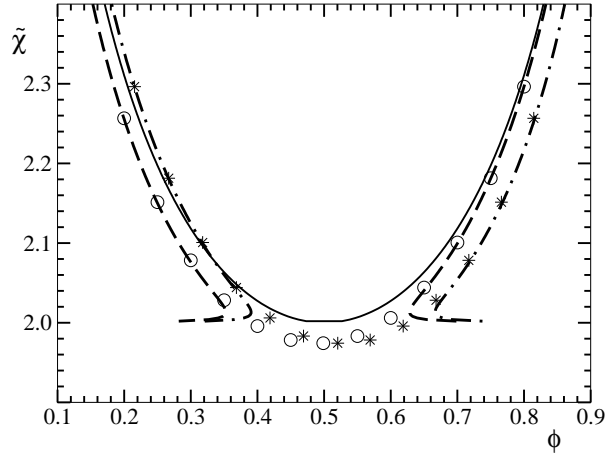}
  \caption{Phase diagram for the chemically-polydisperse Flory-Huggins model
	with a chemically neutral polymeric solvent, in the $(\phi,\ct)$-plane.
	Solid line: monodisperse binodal. Dashed line:
	perturbative cloud curve (8\% polydispersity). Dash-dotted line:
	perturbative shadow curve. Circles: exact cloud point. Stars: exact
	shadow point\protect\cite{Sollichprivate}.}
  \label{FHpd}
  \end{center}
\end{figure}

We consider a Flory-Huggins-style \cite{Flory71} description of a polymer blend.
The reference system is a binary mixture of polymers (species `${\cal A}$' and
`${\cal B}$') of
equal size $L$, but two different chemical types. The chemical nature of
each species is parameterised by a number in the interval $(-1,1)$, which
may be interpreted as the hydrophobicity of the molecule. We make species
${\cal B}$ chemically `neutral' (hydrophobicity zero), while the value for
species ${\cal A}$ is $a$. If the concentration of ${\cal A}$-polymers is
$\phi$ then that of ${\cal B}$-polymers is $1-\phi$, and the mean-field free
energy density is given by
\begin{equation}
\label{FHmono}
	L\F_\m = \phi\ln\phi + (1-\phi)\ln(1-\phi) - \chi\phi_1^2
\end{equation}
where $\phi_1=a\phi$ is the `hydrophobic concentration'. The Flory-Huggins
interaction parameter $\chi$ determines the strength of attraction between
chemically similar species. This model system separates into coexisting
${\cal A}$-rich and ${\cal A}$-poor phases for $\ct\equiv \chi a^2 > 2$, the
concentrations of which are easily found e.g.~from a double-tangent
construction \cite{Callen} on the free energy (Eq.~(\ref{FHmono})). The
resulting phase diagram in the $(\phi,\ct)$-plane is shown by the solid line
in Fig.~\ref{FHpd}.

If component ${\cal A}$ is now made chemically polydisperse, by varying
slightly the constituent monomers on each polymer molecule, while component
${\cal B}$ remains monodisperse and chemically neutral, acting only
as a polymeric solvent, then the overall hydrophobic concentration of polymer
${\cal A}$ is the mean of a distribution:
\[
	\phi_1 = \phi \int a\,p(a)\,\td a = \phi \<a\>
\]
and the free energy density becomes
\begin{equation}
\label{FHpoly}
	L\F = (1-\phi)\ln(1-\phi) - \chi\phi_1^2 +
		\int \phi\,p(a)\ln(\phi\,p(a))\,\td a.
\end{equation}

The model free energy of Eq.~(\ref{FHpoly}) was studied previously to
demonstrate a sophisticated approximation scheme for systematically reducing
the dimensionality of polydisperse phase equilibria problems
\cite{Sollich00}. Accurate cloud (\textsf{o}) and shadow
(\textsf{*}) points were calculated by that method, and are
reproduced in Fig.~\ref{FHpd} (where $\ct\equiv\chi a_0^2$) for a Gaussian
parent with $\<a\>_\p\equiv a_0=0.5$ and polydispersity $\sigma = 8\%$
\cite{Sollichprivate}.

For comparison, our lowest-order perturbative formulae for the cloud and
shadow points (Eqs.~(\ref{cloud},~\ref{shadow})) can be evaluated with ease.
Note that the last term of Eq.~(\ref{FHpoly}) has the form of an ideal free
energy density if we identify $\phi$ with $\rho$ in Eq.~(\ref{Fid}) (up to an
irrelevant term linear in $\rho$). Hence, by writing $a=a_0(1+\epsilon)$ to
expand about the monodisperse reference system with $a=a_0=0.5$,
Eq.~(\ref{FHpoly}) can be cast in the form of Eq.~(\ref{Fex}), with
\begin{eqnarray*}
	A 		&=& -2\ct \phi^2		\\
	B 		&=& 0				\\
	C 		&=& -\ct \phi^2
\end{eqnarray*}
all of which are scalar quantities since only one property (hydrophobicity)
is polydisperse. The cloud and shadow points ($\phi_c=\phi_c^\m+\delta_c$, 
$\phi_s=\phi_s^\m+\delta_s$) are obtained by substituting these values into
Eqs.~(\ref{cloud},~\ref{shadow}) together with the isothermal compressibility
derived from Eq.~(\ref{FHmono}),
$\kappa = (1-\phi)/(\phi-2\ct\phi^2(1-\phi))$, yielding
\begin{eqnarray}
	\phi_c &=& \phi_c^\m + 2\sigma^2\ct^2 \;
	\frac{\phi_c^\m(1-\phi_c^\m) (\phi_c^\m-\phi_s^\m) \phi_s^\m}
	{1-2\ct \phi_c^\m(1-\phi_c^\m)}
\label{FHcloud}	\\
	\phi_s &=& \phi_s^\m + 2\sigma^2\ct^2 \;
   \frac{\phi_s^\m(1-\phi_s^\m) (\phi_s^\m-\phi_c^\m)(2\phi_s^\m-\phi_c^\m)}
	{1-2\ct \phi_s^\m(1-\phi_s^\m)} 
\label{FHshadow}
\end{eqnarray}
where $\phi_c^\m$, $\phi_s^\m$ are the respective points on
the monodisperse binodal (solid line in Fig.~\ref{FHpd}).

Eqs.~(\ref{FHcloud},~\ref{FHshadow}) are plotted in Fig.~\ref{FHpd} as dashed and
dash-dotted lines respectively. As expected, the perturbative expansion
scheme breaks down near the critical point, where the compressibility
diverges and the separate reference phases disappear. This invalid regime
extends over only a small range of $\ct$, and the phase boundaries rapidly
converge, with increasing $\ct$, to a very accurate answer.

Notice that the miscibility gap is
broadened by polydispersity. Equation (\ref{gap}) shows that the polydisperse
perturbation to the gap must be positive whenever $B=B'=0$, i.e.~when there
is no explicit dependence of the excess free energy on the variance of the
distribution of species.

We also have formulae prescribing the chemical properties of the two phases.
By definition, the distribution of polymers in the cloud phase is just the
same as the parent distribution that was input to the system. The shadow
phase, on the other hand, has chosen its own preferred mix. The mean
hydrophobicity of type-${\cal A}$ polymers in this phase is given by
Eq.~(\ref{means}a) (or equivalently Eq.~(\ref{p})) as
$\<a\>_s=a_0(1+\<\epsilon\>_s)=a_0(1+2\sigma^2\ct(\phi_s-\phi_c)
+{\cal O}(\sigma^3))$, which is also in good agreement with exact numerical
data.

\section{Example: Sterically-stabilised colloid-polymer mixtures}
\label{example2}

To further illustrate the utility of the formulae derived in this article,
let us make a case study of a familiar complex fluid: a colloid-polymer
mixture \cite{Ilett95}.
The phase behaviour of the monodisperse system is first briefly
reviewed. A theoretical understanding of its behaviour is well-established
\cite{Lekkerkerker92}.
However, the effects of colloidal polydispersity on the system have not
previously been calculated, despite the fact that the results of colloidal
synthesis are inevitably slightly polydisperse. The effects of polymeric
polydispersity on the system, which have been modelled previously
\cite{Sear97}, will not be analysed.

\subsection{Established monodisperse behaviour}

In this system, small balls (typically of diameter 10~nm to
1~$\mu$m) of PMMA (`Perspex') \cite{Ilett95} or silica \cite{deHek81} are suspended
in an organic liquid. To avoid the balls (or `colloidal latices') clumping due
to van der Waals attraction, they
are each given a short brush-like coat of `steric' polymer chains, making the
latices behave as hard spheres, with no long-range interactions but a very
short-range strong repulsion. Hard spheres of this kind are known
\cite{Russel89} to have a simple phase diagram, and carry no latent heat. The
only controlling parameter is the fraction of space $\phi$ occupied by the spheres.
A single phase transition exists, between an amorphous fluid state and an FCC
crystal, which coexist at $\phi\approx 0.494$ and $\phi\approx 0.545$
respectively \cite{Hoover68}. The phase behaviour is enriched by the addition
of free polymer chains to the mixture. Each of these chains, being very long,
bunches itself into the form of a self-avoiding
random walk. Such a `ball of string' cannot penetrate a colloidal latex, and
so interacts with the colloid particles as if it too were a hard sphere. On the
other hand, when two randomly coiled polymer molecules meet, they {\em can}
interpenetrate and, with a suitable choice of suspending solvent, can be made
almost entirely non-interacting\footnote{In particular, the second virial
coefficient of the polymeric osmotic pressure can be made to vanish, though
higher virial coefficients remain \protect\cite{Berry66}.},
like an ideal gas \cite{Ilett95,Lekkerkerker92}. The ideal gas pressure that
they exert on the surface of an isolated colloidal hard sphere is isotropic.
However, if a second hard sphere is so close to the first that the
non-penetrating polymer coils cannot fit between them, then the polymeric
pressure is not felt on the adjacent parts of the colloids, so they are pushed
together.

Asakura and Oosawa \cite{Asakura54} showed that this mechanism could be
described in terms of an effective attraction between colloidal spheres, known
as the depletion interaction. Thus, rather than considering a mixture of two
components (colloid and polymer) in the solvent, one can imagine a
single-component system, consisting of spheres with a hard core repulsion and a
longer-range
pairwise\footnote{A more accurate description would include corrections to the
pairwise attraction, since three or four latices in close proximity exclude
polymer coils from a region with non-trivial geometry.} attraction. This
effective attractive potential $U(r)$ between two latices of radius $a$ and
centre-to-centre separation $r$ is simply the product of the (ideal) polymeric
osmotic pressure $\Pi_p$
and the volume of space from which the centres of polymer coils are
excluded by the hard spheres \cite{Russel89}. At $r=2a$
(contact of the two hard spheres) the potential rises discontinuously to
infinity, and for $r>2(a+r_g)$, where $r_g$ is the radius of gyration
of a polymer chain, there is no interaction, since the diameter of a
polymer coil ($2r_g$) sets the range of the effective potential.

The phase diagram of this system in the $(\phi,\Pi_p)$-plane has been calculated
for the monodisperse case
\cite{Lekkerkerker92}, using results from the Percus-Yevick integral equation
theory for hard spheres \cite{Lebowitz64}. It was found (in agreement with
experiments \cite{Ilett95}) to resemble the phase diagram of simple atomic
substances such as argon, with $\Pi_p^{-1}$ playing the r\^{o}le of temperature.
As well as the colloidal crystal, there are low- and high-concentration
amorphous phases, dubbed `colloidal gas and liquid', which meet at a
critical point. We shall study the `fluid-fluid' coexistence of these
amorphous phases, thus avoiding problems of non-ergodicity arising in the
polydisperse crystal \cite{Colin}.

\subsection{Calculating the polydisperse phase equilibria}

The strength of the treatment of polydispersity in this article is that it uses,
as input, properties of the monodisperse system. The phase
behaviour of monodisperse colloid-polymer mixtures is well understood, so
application of the new formulae for polydisperse phase equilibria will be
straightforward. In addition to the monodisperse phase diagram, we require
the functions $A(\rho)$ and $B(\rho)$, defined by Eq.~(\ref{Fex}), which are
scalars in this case since the particle radii alone are polydisperse.

Unlike the previous example in section~\ref{FH}, we are not given an
expression for the free energy of the polydisperse system. According to
Eq.~(\ref{ABC}a), $A(\rho)$ can be found from the monodisperse free energy, for
which we use the expression given in Ref.~\cite{Lekkerkerker92}. It is simply
the derivative of the free energy with respect to the radius of monodisperse
colloidal particles (scaled by the radius), at fixed number density. Hence an
expression for $A(\rho)$ is straightforwardly found, so long as care is taken
not to overlook the dependence of concentration on particle size,
$\phi=\frac{4}{3}\pi a^3\rho$, and to keep the polymer size $r_g$ fixed while
taking the derivative.

It is not so trivial to evaluate $B(\rho)$ without a prescribed polydisperse
free energy. Firstly, we must specify the difference between the Hamiltonian of
the system with slightly polydisperse colloid, and that of a monodisperse
reference system. Let us write the Hamiltonian of the colloid-polymer mixture
with polydisperse colloid as a sum of two parts:
\begin{equation}
	H = H_{\rm HS} + H^{\rm eff}_{\rm dep}
\end{equation}
the Hamiltonian $H_{\rm HS}$ of the hard-sphere colloid, plus effective
interactions $H^{\rm eff}_{\rm dep}$ due to the depletion of
polymer. Following the analysis \cite{Lekkerkerker92} of the
monodisperse case, we assume that the resulting free energy is also
separable into two parts:
\begin{equation}
\label{Fcolpol}
	F = F_{\rm HS} + F_{\rm dep}
\end{equation}
where $F_{\rm HS}$ is approximately the free energy of polydisperse hard
spheres in the absence of polymer. This is consistent with the
Weeks-Chandler-Andersen (WCA) approximation \cite{Weeks71} which states that
the spatial distribution of repulsive particles (in this case hard spheres) is
not significantly altered by the introduction of additional interactions which
are purely attractive (here contained in $H^{\rm eff}_{\rm dep}$).
Accordingly, the
polymeric contribution to the free energy $F_{\rm dep}$ will be approximated by
averaging the depletion Hamiltonian $H^{\rm eff}_{\rm dep}$ over the pure hard
sphere distribution.

Note that the structure of Eq.~(\ref{Fcolpol}) implies that the desired
function $B(\rho)$ also splits into two terms,
\begin{equation}
\label{Bsplit}
	B = B_{\rm HS} + B_{\rm dep}
\end{equation}
The depletion interaction is soft, so the thermodynamic perturbation theory
of section~\ref{pert} is suitable for calculating its contribution in
$B_{\rm dep}$.
The hard sphere potential, on the other hand, is non-differentiable and
therefore inappropriate for description in terms of an energetic perturbation,
as noted above. Other methods will be used to deal with the hard sphere part
$B_{\rm HS}$.

Let us tackle the polymeric (depletion) contribution first. As stated above,
the effective Hamiltonian (i.e.~the polymeric free energy for a fixed
configuration of colloidal spheres) $H^{\rm eff}_{\rm dep}$ is simply the
product of the ideal polymeric osmotic pressure and the volume of space from
which centres of polymer coils are excluded. Let us define 
the polymer-colloid size ratio as $\xi\equiv r_g/a$, and
$v_s\equiv\frac{4}{3}\pi a^3$ to be the volume of a reference sphere with the
mean radius. An isolated colloidal sphere of radius $a(1+\epsilon)$ excludes
polymer coils from a volume $v_s(1+\xi+\epsilon)^3$. Hence, even for
well-separated colloids beyond the interaction range, there is a {\em bulk}
contribution to the depletion Hamiltonian,
\begin{equation}
      H_{\rm bulk}=\frac{\hat{\Pi}_p}{\xi^3}\sum_{i=1}^N(1+\xi+\epsilon_i)^3
\end{equation}
where the polymeric osmotic pressure has been re-scaled:
$\hat{\Pi}_p\equiv v_s \xi^3 \Pi_p$, so that the $(\phi, \hat{\Pi}_p)$-plane
will correspond to phase diagrams presented in Ref.~\cite{Lekkerkerker92} for
colloid-polymer mixtures.
For monodisperse colloid, $H_{\rm bulk}$ is usually ignored, since it is a
constant, independent of volume fraction $\phi$. For a polydisperse system, on
the other hand, it cannot be neglected, as it contains
$\epsilon$-dependence. From Eq.~(\ref{B}), its contribution to $B(\rho)$ is
\begin{equation}
\label{Bbulk}
	B_{\rm bulk} = 3 \hat{\Pi}_p \rho \, (1+\xi)/\xi^3.
\end{equation}

When two
colloidal latices, of radii $(1+\epsilon_1)a$ and $(1+\epsilon_2)a$, have
a centre-centre separation
$(2+\epsilon_1+\epsilon_2)a<r<(2+2\xi+\epsilon_1+\epsilon_2)a$, their
individual regions of polymeric exclusion overlap, so that the total volume
available to polymer increases, giving rise to the depletion attraction.
It can be shown by some elementary geometry that the overlap volume of the two
spheres of depletion; radii $r_1=(1+\xi+\epsilon_1)a$ and
$r_2=(1+\xi+\epsilon_2)a$; is
\[
	V_{\rm overlap} = \pi \left[
	\frac{r^3}{12} - \frac{(r_1^2+r_2^2)r}{2} + \frac{2(r_1^3+r_2^3)}{3}
	- \frac{(r_1^2-r_2^2)^2}{4r}	\right].
\]
The resulting interaction potential
$U(r,\epsilon_1,\epsilon_2)=-\Pi_p V_{\rm overlap}$ gives a
pairwise contribution to the effective Hamiltonian, as in Eq.~(\ref{central}).
Its derivatives, as defined in Eqs.~(\ref{potlderivs}), are
\begin{mathletters}
\begin{eqnarray}
\label{depderivs}
	U_1(r) &=& -\frac{3(1+\xi)}{4\xi^3} (2+2\xi-r/a)\: \hat{\Pi}_p	\\
	U_{11}(r) &=& \frac{3}{4\xi^3} \left( 
	\frac{r}{a}-4[1+\xi]+2\frac{a}{r}[1+\xi]^2 \right) \: \hat{\Pi}_p
\end{eqnarray}
which can be substituted into Eq.~(\ref{symmetric}b) for $B_{\rm dep}$.
Additionally, that equation requires the pair and three-point distribution
functions $g_\m(r)$, $g^{(3)}_\m(\vecr,\vecr')$ for the monodisperse system.
At the level of the WCA approximation, these are replaced by the pure
hard-sphere (HS) distribution functions, $g\approx g^{\rm HS}$, for which we
use the Percus-Yevick expression \cite{Wertheim63}. As
a simplifying assumption, let us also use the Kirkwood superposition
approximation \cite{Kirkwood35} for the three-body correlations,
\[
	g^{(3)}(\vecr,\vecr') \approx g(r)\,g(r')\,g(|\vecr-\vecr'|).
\]
\end{mathletters}

It remains only to find the hard-sphere contribution to $B$.
Unfortunately, the methods developed here do not enable a first-principles
derivation of this quantity, so we must look to other analyses of
polydisperse hard spheres for an evaluation of $B_{\rm HS}$. In particular,
we refer to the free energy expression for the polydisperse
hard-sphere fluid due to Boublik, Mansoori, Carnahan, Starling
and Lealand \cite{Boublik70} (BMCSL), which is known to reduce to the
Carnahan-Starling free energy \cite{Carnahan69} in the monodisperse limit. The
application of this expression
perhaps requires some clarification. The use of results from other studies of
polydispersity by no means depreciates the present analysis. In cases where a
polydisperse free energy is already known, one is spared the
application of first-principles methods such as section \ref{evaluation}.
However, the results of section \ref{phaseequilib} (or
equivalent methods) are still required, in order to derive (to lowest-order)
the phase equilibria from that free energy. Indeed, the gap between a
knowledge of the free energy and a knowledge of the phase equilibria is
evidenced by the BMCSL free energy, whose regime of thermodynamic
instability was only recently established \cite{Warren99}.

To extract the required function, the BMCSL free energy is cast in the form
of Eq.~(\ref{Fex}) by expanding to order $\epsilon^2$, at a density
$\rho=\phi_0/v_s$, giving
\begin{eqnarray}
	v_s\,\F_{\rm HS}^\ex &=& \frac{\phi_0^2 (4-3\phi_0)}{(1-\phi_0)^2}
	+ \<\epsilon\> \frac{6\phi_0^2(2-\phi_0)}{(1-\phi_0)^3}
	+ \<\epsilon^2\> 3\phi_0 \left[
	\frac{\phi_0(1+3\phi_0-2\phi_0^2)}{(1-\phi_0)^3}-\ln(1-\phi_0) \right]
	\nonumber \\
 & &	+ \<\epsilon\>^2 3\phi_0 \left[
	\frac{\phi_0(1+2\phi_0)(3+\phi_0-\phi_0^2)}{(1-\phi_0)^4}
	+\ln(1-\phi_0) \right] \quad +\O(\epsilon^3).
\end{eqnarray}
The zeroth-order term is the excess free energy of a monodisperse hard-sphere
fluid, and recovers the Carnahan-Starling equation of state. By comparison
with Eq.~(\ref{Fex}), the coefficient of $\<\epsilon^2\>$ is the required
function $v_s B_{\rm HS}$.

\begin{figure} \displaywidth\columnwidth
  \epsfxsize=8.5cm
  \begin{center}
  \leavevmode\epsffile{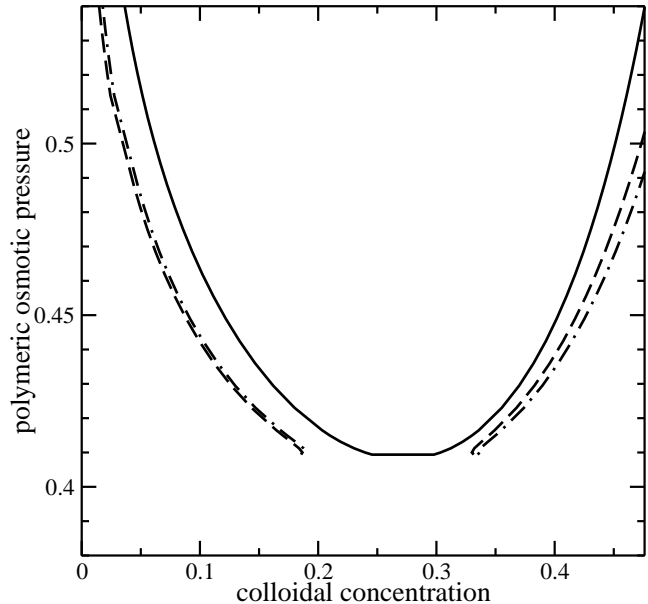}\vspace{3mm}
  \caption{Colloidal gas-liquid phase boundaries for colloid-polymer mixture
	with size ratio $\xi=0.4$, in the plane of colloidal volume fraction
	$\phi$ and polymeric osmotic pressure $\hat{\Pi}_p$. Solid line:
	monodisperse binodal. Dashed/dash-dotted lines: perturbative
	cloud/shadow curves for 8\% polydisperse colloid.}
  \label{PD}
  \end{center}
\end{figure}

For definiteness, let us consider a polymer-colloid size ratio $\xi=0.4$. For
a given polymeric osmotic pressure $\hat{\Pi}_p$, the coexisting concentrations $\phi$
of monodisperse colloidal fluid phases are given in Ref.~\cite{Lekkerkerker92}.
These values for the concentrations of coexisting monodisperse gas
($\phi^g_\m$) and liquid ($\phi^l_\m$) phases are substituted into the
expressions calculated above to obtain values of $A$ and $B$ for the
coexisting phases.
The monodisperse gas-liquid phase boundaries $\phi^g_\m$, $\phi^l_\m$ are
shown as a solid line in Fig.~\ref{PD} from the critical point at
$\hat{\Pi}_p\approx0.41$ to the triple point at $\hat{\Pi}_p\approx0.54$.

The calculated values of $A$ and $B$ and their derivatives are used in
Eqs.~(\ref{cloud},~\ref{shadow}) to find the change in each phase's cloud- and
shadow-point densities due to polydispersity, which are translated into
concentrations via Eq.~(\ref{phitransform}). These shifts in concentration are
added to the monodisperse phase boundary in Fig.~\ref{PD} to give the
resulting cloud and shadow points
\mbox{$\phi^{c,s}=\phi^{c,s}_\m+\delta\phi^{c,s}$} for polydispersity
$\sigma=8\%$, shown as dashed and dash-dotted curves respectively. We see
from the figure that the expansion has remained well controlled (has not
blown up) even for values of $\hat{\Pi}_p$ very close to criticality of the
reference system. The theory predicts a widening of the coexistence region
(the miscibility gap) so that liquid will condense from a polydisperse gas of
lower concentration than in the monodisperse case. We also observe that when
the gas phase exists in an infinitesimal quantity (at the shadow point), it
is more concentrated than at its cloud point. The same applies to the liquid
phases. Similar results are also found for different size ratios $\xi$
(studied for $0.3<\xi<1$).

The above calculation of $B(\rho)$ for the colloid-polymer mixture involves a
degree of inexactitude, since the radial distribution function is
approximated by the Percus-Yevick hard-sphere distribution function, while
three-point correlations are even more crudely estimated. To assess the
importance of precision in the correlations, the cloud and shadow points were
re-calculated using a Heaviside step function $g(r)=\Theta(r-2a)$
in place of the Percus-Yevick radial distribution. The polydispersity-induced
shift in the high-density cloud point changes sign when the
correlations are neglected, indicating, not surprisingly, that the correct
correlations are important in determining the high-density phase boundary.
Though the positions of the other phase boundaries are also affected by the
change in $g(r)$, certain qualitative features remain unaltered, and are
therefore expected to be reliable predictions of our approximate model. In
particular, each shadow point remains more concentrated than the
equivalent cloud point and, far from the critical point, the gases are
shifted to lower concentrations by polydispersity.

\begin{figure} \displaywidth\columnwidth
  \epsfysize=8cm
  \begin{center}
  \leavevmode\epsffile{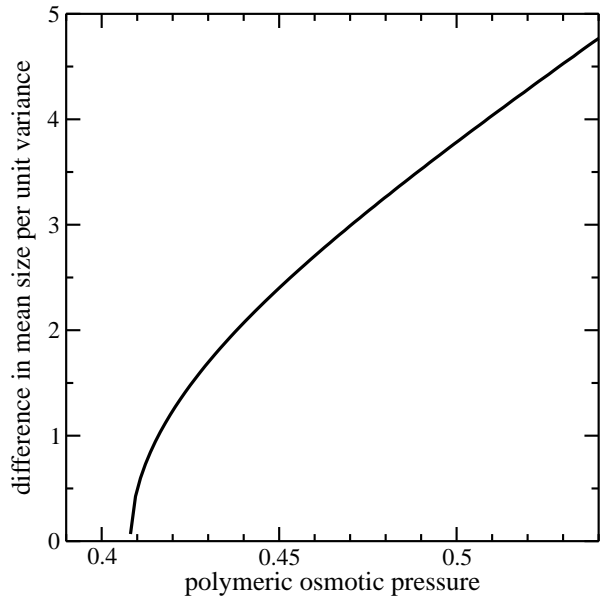}
  \caption{Fractionation per unit variance
	($[\<\epsilon\>]^l_g/\sigma_\p^2=-[A/\rho]^l_g$) against polymeric
	osmotic pressure $\Pi_p$ for gas-liquid coexistence of a
	colloid-polymer mixture with size ratio $\xi=0.4$ and colloidal
	polydispersity $\sigma_\p=6\%$.}
  \label{fractionation}
  \end{center}
\end{figure}

According to Eqs.~(\ref{Deltap}) and (\ref{momentsep}), the difference between
the normalised populations of the coexisting phases is proportional to
$-[A/\rho]^l_g$ and the fractional difference in mean particle size is
$[\<\epsilon\>]^l_g=-[A/\rho]^l_g\,\sigma_\p^2$ where $\sigma_\p$ is the
overall polydispersity. So the coefficient $-[A/\rho]^l_g$ determines the
amount of partitioning or fractionation. It is plotted for the gas-liquid
phase boundary (from $\hat{\Pi}_p=0.41$ to $0.54$) in
Fig.~\ref{fractionation}. Note that it is positive, indicating that, on
average, the liquid phase contains larger particles than the gas. The
hard-sphere contribution is negative, but is outweighed by the depletion part.

\section{Summary}

The results derived here are exact in the limit of low polydispersity, and can
therefore provide accurate information on a great many systems, whose
manufacture attempts to approximate the ideal of monodisperse constituents.
Additionally, the results will be of pragmatic use even in systems with a wide
scatter of particle properties, in providing qualitative predictions and
estimates for the effects of polydispersity. In any case, minimal effort is
required to substitute values into the formulae.

Equations (\ref{p}) and, equivalently, (\ref{Deltap}) show that, in the narrow
limit, the difference (due
to fractionation) between normalised distributions in coexisting phases does
not depend on the volumes of the phases, and is in fact remarkably simple,
requiring only one system-dependent parameter per polydisperse property. From
Eq.~(\ref{momentsep}), the $m$th moment about the centre of the distribution
differs, between coexisting phases, by an amount proportional to the $(m+1)$th
moment of the parent\footnote{plus terms of the order of the $(m+2)$th moment,
which become important in a near-symmetric distribution if $m$ is even
\protect\cite{Xu00}.}. For hard spheres, the constant of proportionality has
been determined (Eq.~(\ref{AHS})), and agrees with data from simulation
\cite{Bolhuispriv}. The calculation is indistinguishable from the data at 2\%
polydispersity, and deviates by only 5\% at 4\% polydispersity.

Additionally, at the nematic cloud point of the polydisperse Zwanzig model of
hard rods \cite{Clarke00}, the form of Eq.~(\ref{momentsep}) has been shown
\cite{Searpriv} to hold to within 5\% up to the remarkably large
polydispersity of 50\%. In section \ref{fractsection}, it was
shown that this simple rule can lead to `convective fractionation' for
multiply-polydisperse systems, whereby one property is partitioned between
phases due to a driving force on another.

The expressions found for the shift in the phase boundaries due to
polydispersity, at a general coexistence (Eq.~(\ref{shift})) and at the cloud
(Eq.~(\ref{cloud})) and shadow points (Eq.~(\ref{shadow})), show that the shift in
the density of a phase is proportional to its isothermal compressibility and to
the square of the polydispersity. This explains why these phase boundaries are
parabolic near $\sigma=0$ when plotted as a function of polydispersity
$\sigma$ in a range of models; see
e.g.~\cite{Bolhuis96,Kofke99,Bartlett99,Bartlett97,Clarke00,Almarza99}. The
other parameters on which these shifts depend can be extracted directly from
the form of the polydisperse free energy if it is known or, if not, from the
Hamiltonian via the thermodynamic perturbation theory in section \ref{pert}. In
the case of pairwise interactions, the relevant parameters are obtained from
a knowledge of only two- and three-body correlations in the reference system,
even when larger clusters are responsible for the existence of the phase
transition in question. This illustrates a strength of the small-polydispersity
expansion: that the underlying reference system takes care of complicated
many-body interactions, so they do not need to be calculated in the analysis of
the polydispersity.

The application of the methods to the fluid-fluid coexistence of
colloid-polymer mixtures in section \ref{example2} revealed that colloidal
polydispersity leads to a widening of the coexistence region, and favours
larger particles living in the liquid phase. Similar methods are used in
appendix \ref{correlations} to find how polydispersity alters correlation
functions.

Other methods exist for calculating polydisperse phase equilibria. Whatever the
theoretical formalism by which one chooses to analyse a polydisperse system,
the results, if correct, will tend in the limit of low polydispersity to
Eq.~(\ref{Deltap}) for the degree of fractionation, and Eq.~(\ref{shift}) for the
movement of the binodals.

\section{Acknowledgements}

Many thanks for informative discussions go to Peter Sollich, David Fairhurst,
Michael Cates, Paul Bartlett, Peter Bolhuis, Richard Sear, Patrick Warren and
Wilson Poon. The work was funded by a Royal Society of Edinburgh/SOEID personal
research fellowship, and EPSRC grant number GR/M29696.

%\end{multicols}

\appendix
\section{Thermal averages and spatial correlations}
\label{correlations}

In polydisperse systems, correlation functions are easier to calculate
than phase boundaries, as there is no partitioning of the sample to
consider. The thermal average of a stochastic quantity $\zeta$ in a
single polydisperse phase is given by the usual perturbative expression
with respect to a monodisperse reference system
\begin{equation}
	\langle\!\langle\zeta\rangle\!\rangle
	= \frac{\langle\!\langle\zeta\,e^{-\hat{H}}\rangle\!\rangle_\m}
	{\langle\!\langle e^{-\hat{H}}\rangle\!\rangle_\m}.
\end{equation}
We shall require the second-order expansion of this expression, in the
small quantity $\hat{H}$, which must therefore be small (in units of $k_{B}T$),
\begin{equation}
\label{expandzeta}
	\langle\!\langle\zeta\rangle\!\rangle
	  = \langle\!\langle\zeta\rangle\!\rangle_\m
	  - \Big(1+\langle\!\langle\hat{H}\rangle\!\rangle_\m\Big)
	  \Big(\langle\!\langle\zeta\hat{H}\rangle\!\rangle_\m
	  -\langle\!\langle\zeta\rangle\!\rangle_\m
	  \langle\!\langle\hat{H}\rangle\!\rangle_\m\Big) \nonumber \\
          +\frac{1}{2}\Big(\langle\!\langle\zeta\hat{H}^2\rangle\!\rangle_\m
	  - \langle\!\langle\zeta\rangle\!\rangle_\m
	  \langle\!\langle\hat{H}^2\rangle\!\rangle_\m\Big) + \O(\hat{H}^2).
\end{equation}
Let us restrict the discussion to a system of particles interacting
via a pairwise-additive, symmetric, isotropic, central potential
$U(r,\ep_1,\ep_2)$, for which the Hamiltonian is given in
Eq.~(\ref{central}). With the Hamiltonian Taylor-expanded to second order in
$\ep$, the perturbation can be written
\[
	\hat{H} = \frac{1}{2} \sum_{i=1}^{N}
	\sum_{\stackrel{\mbox{\scriptsize $j\!\!=\!\!1$}}{\neq i}}^{N}
	\! \Big\{
	(\ep_i+\ep_j) \cdot \dU(|\vecr_i-\vecr_j|)
	+ \;\frac{1}{2} (\ep_i\,\ep_i+\ep_j\ep_j) : \ddU(|\vecr_i-\vecr_j|)
	+ \;\ep_i\,\ep_j : \dcU(|\vecr_i-\vecr_j|) \Big\} + \O(\epsilon^3)
\]
in terms of the derivatives of the interaction potential defined in
Eqs.~(\ref{potlderivs}). Substituting this expression into
Eq.~(\ref{expandzeta}), we keep terms to second order in $\ep$, and
measure the deviations $\ep$ with respect to the mean properties of
particles in the single-phase system, so that $\<\ep\>=\orig$. This
yields
\begin{eqnarray}
	\<\!\<\zeta\>\!\> &=& \<\!\<\zeta\>\!\>_\m 
	- \<\ep\ep\> : \frac{1}{2}\sum_{i,j}\Big\{
	\<\!\<\zeta \; \uuPsi(|\vecr_i-\vecr_j|)\>\!\>_\m
	- \<\!\<\zeta\>\!\>_\m 
	\<\!\< \uuPsi(|\vecr_i-\vecr_j|) \>\!\>_\m \Big\}  \nonumber \\
& &	+ \<\ep\ep\> : \frac{1}{2} \sum_{i=1}^N
        \sum_{\stackrel{\mbox{\scriptsize $j\!\!=\!\!1$}}{\neq i}}^{N}
        \sum_{\stackrel{\mbox{\scriptsize $k\!\!=\!\!1$}}{\neq i,j}}^{N}
	\Big\{
	\<\!\<\zeta\;\dU(|\vecr_i-\vecr_j|)\dU(|\vecr_i-\vecr_k|)\>\!\>_\m
	- \<\!\<\zeta\>\!\>_\m
	\<\!\<\dU(|\vecr_i-\vecr_j|)\dU(|\vecr_i-\vecr_k|)\>\!\>_\m
	\Big\}
\label{zetaeq}
\end{eqnarray}
where
\[
	\uuPsi(r) \equiv \ddU(r) - \frac{1}{2} \dU(r)\dU(r).
\]
The thermal averages of functions of the stochastic particle positions
can be re-written in Eq.~(\ref{zetaeq}) in terms of the number density field in
the monodisperse system, which at any instant is
$\rho(\vecr)=\sum_{i=1}^{N}\delta^{(3)}(\vecr-\vecr_i)$, where
$\delta^{(3)}$ is the three-dimensional Dirac delta function.
Finally, the thermal average of a given stochastic variable $\zeta$ becomes
\begin{eqnarray}
	\<\!\<\zeta\>\!\> &=& \<\!\<\zeta\>\!\>_\m - \frac{1}{2}\<\ep\ep\> :
	\int \td\vecr'\,\td\vecr''\; \Big(
	\<\!\<\zeta\,\rho(\vecr')\rho(\vecr'')\>\!\>_\m
	- \<\!\<\zeta\>\!\>_\m \<\!\<\rho(\vecr')\rho(\vecr'')\>\!\>_\m
	\Big) \, \uuPsi(|\vecr'-\vecr''|)	\nonumber  \\
& &	+ \frac{1}{2} \<\ep\ep\> : \int \td\vecr'\,\td\vecr''\,\td\vecr'''\;
	\Big( \<\!\<\zeta\,\rho(\vecr')\rho(\vecr'')\rho(\vecr''')\>\!\>_\m
        - \<\!\<\zeta\>\!\>_\m 
	\<\!\<\rho(\vecr')\rho(\vecr'')\rho(\vecr''')\>\!\>_\m
	\Big) \; \dU(|\vecr'-\vecr''|)\,\dU(|\vecr'-\vecr'''|).
\label{zeta}
\end{eqnarray}

A case of interest, for instance, is $\zeta=\rho(\orig)\,\rho(\vecr)$.
Then Eq.~(\ref{zeta}) gives the perturbation, due to polydispersity, of
the pair correlation function for total density as
\begin{eqnarray}
&&	\<\!\<\rho(\orig)\rho(\vecr)\>\!\> - 
	\<\!\<\rho(\orig)\rho(\vecr)\>\!\>_\m
	= \; - \frac{1}{2}\<\ep\ep\> :
	\int \td\vecr'\,\td\vecr''\; \Big(
        \<\!\<\rho(\orig)\rho(\vecr)\rho(\vecr')\rho(\vecr'')\>\!\>_\m
        - \<\!\<\rho(\orig)\rho(\vecr)\>\!\>_\m 
	\<\!\<\rho(\vecr')\rho(\vecr'')\>\!\>_\m
        \Big) \, \uuPsi(|\vecr'-\vecr''|)       \nonumber  \\
&&     + \frac{1}{2} \<\ep\ep\> : \!\!\!
	\int \!\! \td\vecr'\,\td\vecr''\,\td\vecr'''\; \Big(
  \<\!\<\rho(\orig)\rho(\vecr)\rho(\vecr')\rho(\vecr'')\rho(\vecr''')\>\!\>_\m
        \!- \<\!\<\rho(\orig)\rho(\vecr)\>\!\>_\m
        \<\!\<\rho(\vecr')\rho(\vecr'')\rho(\vecr''')\>\!\>_\m
        \Big) \; \dU(|\vecr'-\vecr''|)\,\dU(|\vecr'-\vecr'''|)	\nonumber \\
&&
\end{eqnarray}
which depends, to second order in the standard deviation of species, on
four- and five-point correlations in the monodisperse reference phase.


\begin{thebibliography}{999}
\bibitem{note}  In this case, the `external' fields may be provided by an
admixture of other conserved species, in whose presence our
slightly-polydisperse component exists.
\bibitem{mistake} In \protect\cite{Evans99}, $A/\rho$ for hard spheres was
evaluated mistakenly as $12P^\ex V^{\rm eff}$, where $V^{\rm eff}$ is some
effective volume. The correct expression is $12PV^{\rm eff}$, since the excess
free energy in expanding a sphere is the work done against the total pressure
of the fluid, not just the excess part. Up to second virial coefficient,
$V^{\rm eff}$ was shown to be the volume of a sphere, which is consistent with
the expression $V^{\rm eff}=(\rho^{-1}-P^{-1})/4$ found by comparison with the
exact calculation here.
\end{thebibliography}
\end{document}